\documentclass[twoside]{article}
\usepackage{amsfonts,amssymb,amsbsy,textcomp,marvosym,picins,amsmath,caption,threeparttable,amsthm,subfigure}
\usepackage{eurosym,mathrsfs,fancyhdr,CJK,multicol,graphics,indentfirst,color,bm,upgreek,booktabs,graphicx}
\usepackage{enumitem}
\usepackage{tabularx}
\usepackage[noadjust]{cite}
\usepackage{pifont}
\usepackage{endnotes}
\usepackage{perpage}
\usepackage[symbol*]{footmisc}
\DefineFNsymbols{circled}{{\ding{192}}{\ding{193}}{\ding{194}}{\ding{195}}{\ding{196}}{\ding{197}}{\ding{198}}{\ding{199}}{\ding{200}}{\ding{201}}}
\setfnsymbol{circled}

\newcommand{\upcite}[1]{\textsuperscript{\cite{#1}}}
\newcommand{\hl}[1]{{\color{black}#1}}

\def\footnoterule{\kern 1mm \hrule width 10cm \kern 2mm}

\allowdisplaybreaks
\catcode`@=11
\def\title#1{\vspace{3mm}\begin{flushleft}\vglue-.1cm\Large\bf\boldmath\protect\baselineskip=18pt plus.2pt minus.1pt #1
\end{flushleft}\vspace{1mm} }
\def\author#1{\begin{flushleft}\normalsize #1\end{flushleft}\vspace*{-4pt} \vspace{3mm}}
\def\address#1#2{\begin{flushleft}\vglue-.35cm${}^{#1}$\small\it #2\vglue-.35cm\end{flushleft}\vspace{-2mm}\par}

\catcode`@=11
\def\section{\@startsection{section}{1}{\z@}%
 {-3ex \@plus -.3ex \@minus -.2ex}%
 {2.2ex \@plus.2ex}%
{\normalfont\normalsize\protect\baselineskip=14.5pt plus.2pt minus.2pt\bfseries}}
\def\subsection{\@startsection{subsection}{2}{\z@}%
 {-3ex\@plus -.2ex \@minus -.2ex}%
 {2ex \@plus.2ex}%
{\normalfont\normalsize\protect\baselineskip=12.5pt plus.2pt minus.2pt\bfseries}}
\def\subsubsection{\@startsection{subsubsection}{3}{\z@}%
 {-2.2ex\@plus -.21ex \@minus -.2ex}%
 {1.4ex \@plus.2ex}
{\normalfont\normalsize\protect\baselineskip=12pt plus.2pt minus.2pt\sl}}

\begin{document}
\begin{CJK*}{GBK}{song}
\thispagestyle{empty}
\vspace*{-13mm}
\noindent {\small Gui CY, Zheng L, He BS {\it et al.} A survey on graph processing accelerators: Challenges and opportunities.
JOURNAL OF COMPUTER SCIENCE AND TECHNOLOGY 
}
\vspace*{2mm}

\title{A Survey on Graph Processing Accelerators: Challenges and Opportunities}

\author{Chuang-Yi Gui$^{1}$, \textit{Student Member, CCF}, Long Zheng$^{1,*}$, \textit{Member, CCF, ACM, IEEE} 
\\Bing-Sheng He$^{3}$, \textit{ Senior Member, IEEE, Member, ACM}, Cheng Liu$^{2,3}$, Xin-Yu Chen$^{3}$
\\Xiao-Fei Liao$^{1}$, \textit{Senior Member, CCF, Member, IEEE} and Hai Jin$^{1}$, \textit{Fellow, CCF, IEEE, Member, ACM}}

\address{1}{\hl{National Engineering Research Center for Big Data Technology and System/Services Computing Technology and System Lab/Cluster and Grid Computing Lab, School of Computer Science and Technology, Huazhong University of Science and Technology, Wuhan, 430074, China}}
\address{2}{Institute of Computing Technology, Chinese Academy of Sciences, Beijing, 100190, China}
\address{3}{School of Computing, National University of Singapore, 117418, Singapore}

\vspace{2mm}

\noindent E-mail: \hl{{\{chygui, longzh\}}@hust.edu.cn; hebs@comp.nus.edu.sg;  liucheng@ict.ac.cn; xinyuc@comp.nus.edu.sg; {\{xfliao, hjin\}}@hust.edu.cn}\\[-1mm]

\noindent Received July xx, 2018; revised January xx, 2019.\\[1mm]

\footnotetext{{}\\[-4mm]\indent\quad Survey Paper\\[.5mm]
\indent\quad This work is supported by the National Key Research and Development Program of China under Grant No.~2018YFB1003502, National Natural Science Foundation of China under Grant Nos.~61825202, 61832006, 61628204 and 61702201, and China Postdoctoral Science Foundation Grant Nos.~2018T110765 and 2018M630862. \\[.5mm]
\indent\quad$^*$Corresponding Author
\\[.5mm]\indent\quad \copyright 2019 Springer Science\,+\,Business Media, LLC \& Science Press, China}

\noindent {\small\bf Abstract} \quad  {\small Graph is a well known data structure to represent the associated relationships in a variety of applications, e.g., data science and machine learning. Despite a wealth of existing efforts on developing graph processing systems for improving the performance and/or energy efficiency on traditional architectures, dedicated hardware solutions, also referred to as graph processing accelerators, are essential and emerging to provide the benefits significantly beyond those pure software solutions can offer. In this paper, we conduct a systematical survey regarding the design and implementation of graph processing accelerator. Specifically, we review the relevant techniques in three core components toward a graph processing accelerator: preprocessing, parallel graph computation and runtime scheduling. We also examine the benchmarks and results in existing studies for evaluating a graph processing accelerator. Interestingly, we find that there is not an absolute winner for all three aspects in graph acceleration due to the diverse characteristics of graph processing and complexity of hardware configurations. We finially present to discuss several challenges in details, and to further  explore the opportunities for the future research.}

\vspace*{3mm}


\vspace*{4mm}

\end{CJK*}
\baselineskip=18pt plus.2pt minus.2pt
\parskip=0pt plus.2pt minus0.2pt
\begin{multicols}{2}

\section{Introduction}
For a wide variety of applications, e.g., date science, machine learning, social networks, roadmap and genomics, graph is expressive to represent \hl{the} inherent relationships between different entities. Therefore, graph processing has become a hot topic for solving many real-world problems in both academia and industry. With the growing development of Internet of Things and cloud computing, the size and complexity of graphs are still expanding. This poses great challenges for modern graph processing eco-systems in both performance and energy efficiency. 

There are a large number of studies that attempt to use software solutions to improve the performance and energy efficiency of graph processing. From distributed computing environment\upcite{Pregel, GraphLab2012}, to single high-end server\upcite{Ligra}, to the commodity personal computer\upcite{GraphChi, X-Stream}, these systems basically make tremendous efforts on software optimizations for programmability, high performance and scalability under traditional architectures. In an effort to accelerate graph workloads, multi-core CPUs and GPUs have been recently adopted to expose a high degree of parallelism for high perfromance graph iteration, e.g., Medusa\upcite{Medusa}, Cusha\upcite{Cusha}, GunRock\upcite{GunRock}, Frog\upcite{Frog}, MapGraph\upcite{MapGraph} and Enterprise\upcite{Enterprise}. 

Despite a large number of software solutions, \hl{the potentials of graph processing on performance and energy efficiency are still bounded to current hardware architectures}. Real-world graphs often follow a power-law distribution in the sense that most of vertices are associated with a few edges, leading to the fact that prohibitive memory access overhead and low efficiency have occurred on general-purpose processors\upcite{Beamer-Locality, Malicevic-Everything, 8-GRAPHPIM,40-YaoPACT}. The irregularity in graph processing inherently fall short in exploiting memory- and instruction-level parallelism on traditional processors. It is also observed in the previous studies that a wealth of memory bandwidth is actually under-utilized for graph processing on existing commodity multi-core architectures\upcite{40-YaoPACT, 1-GRAPHICIONADO, GraphBIG, Satish-Navigating}. 

Though GPUs have demonstrated compelling  performance on graph processing\upcite{ Medusa,Cusha, GunRock,Groute}, they still suffer from key issues in terms of control and memory divergence, load imbalance and superfluous global memory accesses. More important is that CPUs and GPUs are known for relatively high energy consumption. With the end of Moore's law, \hl{using pure software solutions on traditional architectures is often extremely-difficult to fill the significant gap between the general-purpose architectures and the graph-specific computation for seeking the top performance of graph processing}.

\vspace{2mm}

\begin{center}
\includegraphics[width = 8.33cm]{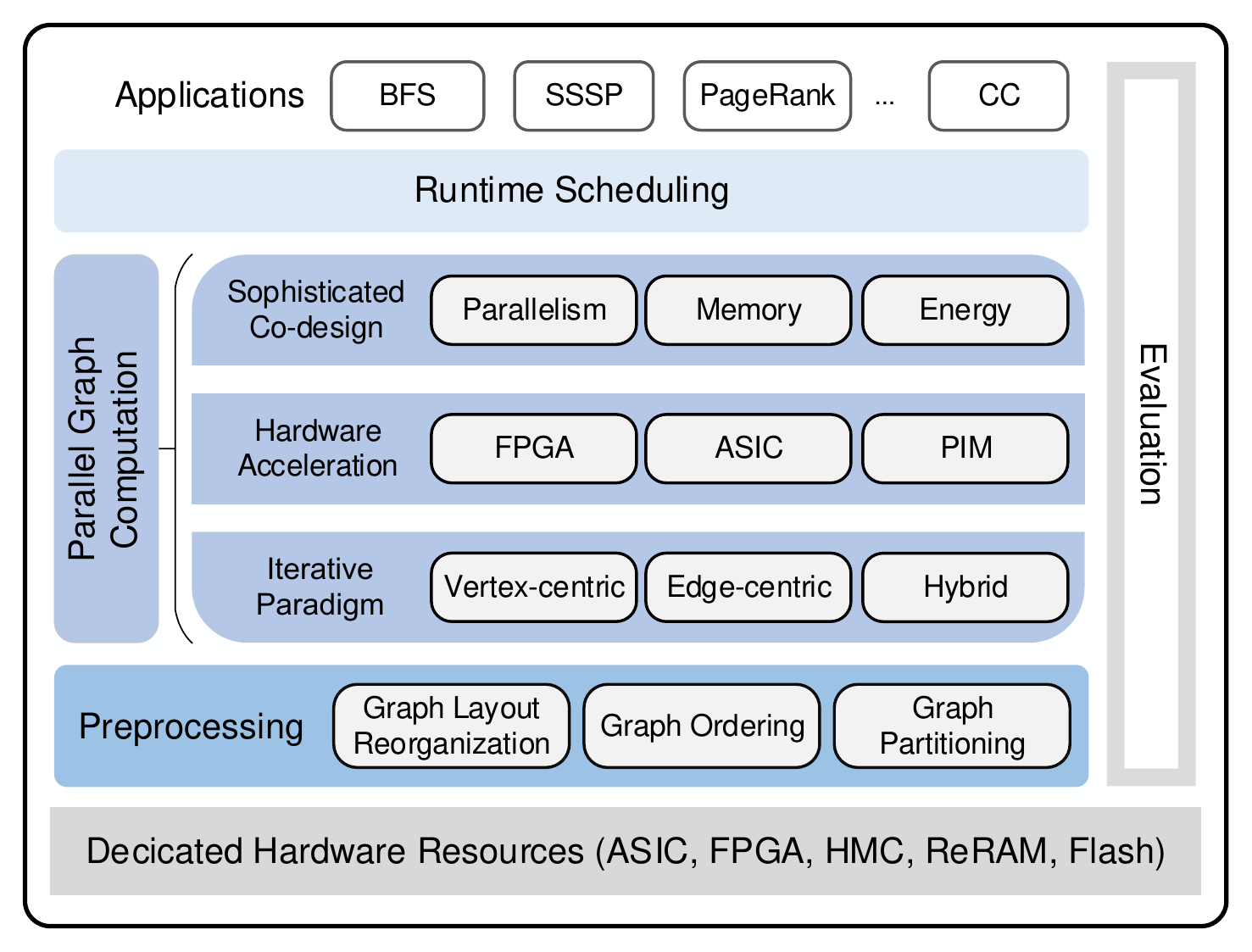}\\
\vspace{2mm}
\parbox[c]{8.3cm}{\footnotesize{Fig.1.~}  Building blocks for graph processing accelerators (with three major aspects: preprocessing, parallel graph computation and runtime scheduling)}
\end{center}

\vspace{1mm}

For graph processing, architectural innovation is imperative. Hennessy and Patterson have also identified the importance, trend and opportunities of \hl{Domain-specific Architecture (DSA)} in their recent technical report\upcite{DSA}. It is pointed out that open sourced architectural implementations\footnote{http://www.riscv.org, Jan. 2019.} 
are the key for the innovations on hardware design\upcite{DDFCP}. The agile chip development can also shorten the development cycle for DSA prototypes\upcite{AGILE}. These guidelines provide one of most effective means for driving the rapid development of graph processing-specific accelerators.  
At this point, hardware platform templates, e.g., \hl{Field Programmable Gate Array (FPGA)} and  \hl{Application-specific Integrated Circuit (AISC)}, are in line with the demand of the times. A large number of industries have already deployed their services on these beneficial hardware platforms for top performance and energy efficiency. For instance, FPGAs have been used in Microsoft datacenter for energy efficiency improvement\upcite{Microsoft-Azure}.

Specifically in terms of graph processing, it has been also witnessed that a large number of relevant studies build their graph processing accelerators based on FPGA\upcite{14-GRAPHSTEP, 20-CYGRAPH,26-FPGP,29-Zhou-Highthroughput,30-ForeGraph} and ASIC\upcite{1-GRAPHICIONADO,2-EAA,4-TUNAO,6-TEMPLATEGAA}. Evaluation on these accelerators has also demonstrated the efficiency and effectiveness of DSA design\upcite{1-GRAPHICIONADO,30-ForeGraph,7-TESSERACT}. 

It is time to review the past and present of graph processing accelerators, and further look into their future development. In this paper, we conduct a systematic review on graph processing accelerators. It aims at exploring the key issues in the design and implementation of graph processing accelerators. As summarized in Fig.1, we have identified a complete set of core components for graph processing accelerator, which involves three major aspects:  preprocessing, graph parallel computation \hl{and} runtime scheduling.

\begin{itemize}[leftmargin=*]
\item \textit{Preprocessing.} Graph processing accelerator often has the limited storage resources. Graphs are needed to be partitioned. Preprocessing is an important component that operates on graph data for trying to make graph dataset fit into the memory capacity of graph accelerator. It is also the key to match a certain processing model and appropriate graph representation before the formal processing.

\item \textit{Parallel Graph Computation.}
Parallel graph computation component serves as the main execution part of graph processing accelerator design. Iterative paradigm is often chosen to define a basic execution pattern for graph iteration that will be mapped to a pipelined hardware circuit. The implementation of this part generally relies on some hardware platform, e.g., FPGA, ASIC, and  Processing-In-Memory (PIM). Different specifications have different concerns on hardware designs and sophisticated software co-designs for high throughput and energy efficiency.

\item \textit{Runtime Scheduling.}
This part aims at how to schedule a large number of graph computational operations on a finite set of hardware resources of graph processing accelerators. The basic metrics for runtime scheduling are to guarantee the correctness and efficiency of graph iteration. The runtime scheduling component often involves data communication, execution mode and scheduling scheme. 
\end{itemize}

Based on aforementioned three aspects, we carefully examine the benchmarks and results \hl{of} existing studies. We find that there is not a clear winner for all these aspects in graph acceleration because of the diverse characteristics of graph processing and the complexity of hardware configureations. We therefore \hl{present and discuss} several challenges in details, and to further explore the opportunities for the future research. One of the major challenges in the existing graph processing accelerators is that the programmability is an important issue for users to express their graph applications. Existing graph processing accelerators typically require labor-intensive efforts for hardware level modifications. 

Great challenges come with great opportunities. Widespread graph applications have a strong demand for energy-efficient graph processing accelerators. Emerging memory devices, e.g., \hl{Hybrid Memory Cube (HMC)\upcite{HMC}, High Bandwidth Memory (HBM)\upcite{HBM}, Resistive Random Access Memory (ReRAM)\upcite{ReRAM}} along with new processing devices, provide us with great opportunities to explore new schemes for graph processing. We believe that this survey summarizes those challenges and opportunities, which can help realize the accelerators with novel hardware-software co-designs.  

The rest of this paper is organized as follows: Section~2 includes an introduction to basic components of graph processing, and briefly summarizes the recent progress on CPUs and GPUs. Section 3 presents some considerations in preprocessing phase. Design and implementation of parallel graph computation are reviewed in Section 4. Section 5 describes the runtime and scheduler part of graph accelerators. Emerging graph accelerators are reviewed and compared in Section 6.  Challenges and opportunities are given in Section 7. Finally, we conclude our work in Section 8.

\section{Preliminaries}
In this section, we first give a brief introduction to the preliminaries of graph processing, including graph representation and several common graph algorithms. Next, we summarize some unique characteristics of graph processing, followed by the related work of graph processing on commodity general-purpose processors. \hl{The characteristics of graph processing and the related work further motivate our survey work on graph processing accelerators}.

\subsection{Graph Representation}
Graph is a data structure consisting of vertices that are further associated with edges. A graph can be typically defined as $G = (V, E)$ where $V$ represents the vertex set and $E$ indicates the edge set. For a directed graph, an edge can be represented as $e=(v_i, v_j)$, indicating that there is an edge pointing from $v_i$ to $v_j$. In particular, vertex and edge can be also attributed with a single or multiple attributes.
Real-world natural graphs, e.g., social networks, usually have the following three common features: 
\begin{itemize}[leftmargin=*]
\item  \textit{Sparsity}.  The average number of vertex degrees is relatively small. The sparsity of graphs can result in poor locality for data accesses.
\item \textit{Power-law Distribution}. A few vertices have associated most of the edges. This can lead to severe workload imbalance issue with a large number of date conflicts when high-degree vertices are being updated.
\item \textit{Small-world Structure}. \hl{Two arbitrary} vertices in the graph can be connected with only a small number of hops. The small-world feature will make it difficult for partitioning the graph efficiently (as will be discussed in \hl{Subsection~3.3}).
\end{itemize}

\subsection{Graph Algorithms}
We review several common graph algorithms with different requirements in computation, communication and memory access. These graph algorithms are also widely studied for the exprimental evaluation in the previous studies\upcite{ Beamer-Locality, Malicevic-Everything,GraphBIG}.


\hl{Breadth-First Search (BFS)} is a basic graph traversal algorithm, which is used as the kernel of Graph500 benchmarks. The neighboring vertices are iteratively accessed from the root vertex until all vertices of the graph are visited. 

\hl{Single Source Shortest Path (SSSP)} is another graph traversal algorithm that computes the shortest paths from a source vertex to other vertices. Different from BFS, it has less number \hl{of redundant computations} in checking edges. Each vertex may be activated more than once. Therefore, it needs more memory space than BFS.

\hl{Betweenness Centrality (BC)} is widely used to measure the importance of a vertex in a graph. The betweenness centrality value of a vertex is calculated by the ratio of shortest paths between any other two vertices. BC algorithm requires to compute the shortest paths between all pairs of vertices.

\hl{PageRank} is one of the most popular algorithms, which calculates the scores of websites\hl{\upcite{PageRank}}. It maintains a PageRank value for each vertex. All the vertices are activated in each iteration. It often needs large memory bandwidth and float point computing ability.

\hl{Connected Components (CC)} is widely used in image regions analysis and clustering applications. Each vertex maintains a label. If vertices are in the same connected region, their labels are set to the same. The algorithm updates the labels of all vertices iteratively until converged.

\hl{Triangle Counting (TC)} is used to measure the number of triangle cliques in the graphs. Each vertex maintains a list of neighbors, and iteratively checks if there are shared neighbors between each connected vertices pair. Number of triangles is calculated by the overlaps.

\hl{Graph Coloring (GC)} is to assign colors to the vertices of a graph so that any two adjacent vertices have different colors. GC can be used in many areas, e.g., traffic scheduling, register allocation during compiling and pattern matching. Basic GC algorithm iteratively colors an active vertex with the color that has not been assigned on any of its neighbours.

\hl{Collaborative Filtering (CF)} is an important machine learning algorithm used for recommendation. Given a bipartite graph where edge values represent the ratings and vertices correspond to the users and \hl{items, CF} runs iteratively on the bipartite graph to find latent features for each vertex, with all the vertices active in each iteration.

\hl{K-core Decomposition (kCore)} is widely used for structure analytics for large cloud networks. This algorithm iteratively removes all the vertices with degrees less than \hl{\it k} such that \hl{\it k}-core subgraphs in each all vertices have degree at least \hl{\it k} are build. 

\hl{Minimal Spanning Tree (MST)} extracts a tree containing all the vertices from an edge-weighted graph with minimum weight. MST is popular in cable network construction, cluster analysis and circuit design. Prim's greedy MST algorithm iteratively chooses the minimum weight edge between vertices in and out of the spanning tree to construct the MST.

\subsection{Unique Features of Graph Processing}
As discussed previously, real-world graphs have the ``power-law" distribution and ``small-world" feature. Besides, graph algorithms differ in computational and memory access requirements. Graph processing generally manifests the unique features as follow.

\begin{itemize}[leftmargin=*]
\item \textit{Intensive Data Access}. On the one hand, graph applications usually lead to a large number of data access requests. On the other hand, graph processing has a high data-access-to-computation ratio. That is, most of the operations in graph processing are related to data accesses.

\item \textit{Irregular Computation}. Due to the power-law distribution, computation workloads for different vertices may vary in a large scale. This will cause severe workload imbalance issue and communication overhead.

\item \textit{Poor Locality}. Data accesses of graph processing are usually random because each vertex may connect to any other random vertices. This feature often leads to heavy overhead of memory accesses.

\item \textit{High Data Dependency}. The data dependency is caused by the nature of connections of vertices in graph. Heavy dependencies make it difficult to explore the parallelism in graph processing. This may cause frequent data conflicts.
\end{itemize}

\subsection{\hl{Brief Introduction} to Graph Processing on Modern Commodity Processors}
Many graph processing systems have been explored on modern commodity general-purpose processors, e.g., CPUs and GPUs. We briefly introduce the related work to motivate our study, and refer readers to recent surveys for more details\upcite{survey-tlav, survey-graphacm, survey-shigpu}.

{\it Graph Processing on CPUs.}\quad
There is a large amount of work that aims at building an efficient system for graph applications on CPUs. Basically, they can be divided into two categories. The first kind is the distributed systems\upcite{PowerGraph,Giraph,GraphX,Arabesque,Powerlyra,Gemini}, which leverage the clusters to support massive graph data. However, this usually suffers from communication overhead, synchronization overhead, fault tolerance and load imbalance issues\upcite{Mizan,Randles-Comparative,LightGraph,Wang-Replication}. Emerging servers can hold most of the graph data in the large main memory. Thus, there \hl{is an amount of} work that exploits the potential of single machine\upcite{Ligra,Galois,GraphMat,zhang-NumaAware}. There are also many disk-based graph processing systems\upcite{ GraphChi, X-Stream,TurboGraph,PathGraph,Gridgraph,NXGraph} which can avoid parts of the challenges in the distributed systems. Recently, Many Integrated Core (MIC) architecture based processors are also explored to improve the performance and efficiency of graph processing\upcite{Mosaic}.

{\it Graph Processing on GPUs.}\quad
GPU is adopted to pursue high performance of graph processing due to its data parallel capability. A number of graph processing systems with GPUs\upcite{Medusa,Cusha,GunRock,Gstream} have been proposed for high-performance graph processing. 
Enterprise\upcite{Enterprise} is developed
to accelerate the performance for BFS algorithm only. There \hl{is also plenty of} work on accelerating CC algorithm\upcite{Soman-Fast}, BC algorithm\upcite{McLaughlin-ScalableBC,Sariyuce-BetweennesBC} and SSSP algorithms\upcite{Davidson-WorkefficientGPUsssp}. Domain-specific graph processing frameworks have been presented to provide high efficiency for the development on GPUs\upcite{Green-Marl}. To support large-scale of graphs, \hl{hybrid} CPU-GPU systems\upcite{Totem,GGraph}, multi-GPUs systems\upcite{ Groute,iBFS} and out-of-memory systems\upcite{GraphReduce,GTS} have been proposed.

{\bf Remarks.}\quad
Despite a significant amount of effort in improving the graph processing performance on general-purpose processors, e.g., CPUs and GPUs, existing graph systems are still far from ideal to exploit the hardware potential of general-purpose processors\upcite{1-GRAPHICIONADO, 40-YaoPACT}. This is due to a significant gap between the general-purpose architectures and the unique features of graph processing. \hl{The graph} processing accelerator is necessary as an alternative approach that might be able to fill this gap. 

Nevertheless, existing studies on CPUs and GPUs have a wealth of experiences in designing graph accelerators (as discussed in the previous studies\upcite{30-ForeGraph, 2-EAA, 4-TUNAO, 7-TESSERACT}). Various kinds of software graph processing models have been proposed to effectively express graph applications in a generic framework. Partitioning methods, out-of-memory processing and hybrid architectures schemes have been explored to support large-scale graphs. 

We next illustrate three aspects of core components of graph accelerators, including {preprocessing}, {parallel graph computation} and {runtime scheduling}. 

\section{Graph Preprocessing}
\hl{The data size of real-world graphs} can easily exceed the on-chip/board memory capacity of graph processing accelerators which is a significant challenge for accelerators. This issue can cause large amounts of I/O and communication cost. In order to make data access efficient, preprocessing of graph data is often required to adapt the data structure onto the target graph accelerators. In this section, we will review the following
major graph preprocessing methods used in the designs of graph processing accelerators.

\begin{itemize}[leftmargin=*]
\item  \textit{Graph Layout Reorganization}.  Graph layout is an important factor to affect the graph processing efficiency. Most previous studies have attempted to reorganize the layout to improve data accessing efficiency from many distinct aspects, e.g., data locality, memory storage, and memory access patterns. 

\item \textit{Graph Ordering}.  Graph ordering aims to change the order of the vertices or the edges, such that data locality with less data conflicts 
can be obtained while the structure of the graph remains the same\upcite{29-Zhou-Highthroughput,9-RPBFS17}.

\item \textit{Graph Partitioning}. Graph partitioning is to divide a large graph into multiple disjoint small subgraphs. 
It usually allows parallel processing of the sub graphs. The processing on 
each sub graph has most of data accesses on the corresponding graph 
partition. This is particularly useful for improving the cache locality or when the memory of the accelerator cannot hold the entire graph. 
\end{itemize}

\subsection{Graph Layout Reorganization}
We will introduce the baseline graph layouts first. There are generally 
two widely-used categories of baseline graph layouts\hl{, i.e.,} edge array and compressed 
adjacency list. In graphs based on the edge array, each element of the array contains 
a pair of integers\hl{, i.e.,} source vertex index and destination vertex index. 
It is convenient to read the edges sequentially from memory. The edge array layout remains widely used in many graph processing systems, especially 
for the edge-centric processing systems. Another improved edge array layout is Coordinate List (COO). It has been widely adopted in graph accelerators\upcite{29-Zhou-Highthroughput,30-ForeGraph, 10-GRAPHR}. 
It has the edge attributes that are stored along with the edges. 

Compressed adjacency list graph originates 
from the adjacency matrix. It typically uses three arrays to store the 
graphs\hl{, i.e.,} the vertex property array of the graph, the edge array with the edges' outgoing/incoming vertex indices only, and the edge array starting indices 
of each vertex in the graph. Suppose outgoing edges are used in the edge array, 
\hl{We name this adjacency list format Compressed Sparse Row (CSR)}. If incoming edges are used 
in the edge array, this layout is called Compressed 
Sparse Column (CSC). The compressed adjacency list graph is relatively compact and 
beneficial to many graph accelerators\upcite{2-EAA,32-FPGAHMCBFS}. 
Note that the edges of each vertex are stored sequentially.

Based on the baseline graph layouts, we have also many novel methods to compress the data size and optimize memory access further.

{\it Combining Information.}\quad
Existing work tends to combine multiple information in the same file of graph data layout so that the data locality can be optimized, and random memory access can be reduced.

For instance, \cite{28-GraphOps} proposes to associate the 
destination vertex property with the edge information such that the vertex 
property can be sequentially accessed to edges with a good locality. Authors in \cite{20-CYGRAPH} 
opt to modify the row pointer array representation in a typical CSR format. They combine the 
vertex status (1 bit for BFS only) and the vertex's neighboring information in an element of the array. 
This method improves the memory access efficiency significantly. 

{\it Encoding Index.}\quad
Using \hl{an} encoding method can compress the graph layout to a small size. Thus, large graphs can be processed on a single accelerator. This is usually done for the index of vertices and edges. 

For example, GraphH\upcite{13-GRAPHH} proposes to squeeze the blank vertex indices 
by re-indexing the vertices of the graph when the number of vertices is smaller than the maximum vertex 
index. The index can also be compressed by grouping them 
with a coarsen id and using less bits to represent the same graph as presented in \cite{1-GRAPHICIONADO,30-ForeGraph}. 
It is also possible to reduce the edge information with \hl{frequency-based} encoding\upcite{36-DegreeBFS}.

{\bf Remarks.}
The baseline graph layouts are useful towards graph accelerators, but they can still be improved for different memory system designs in hardware accelerators. We still have the potential to explore the graph layouts at the aspects of data locality, memory access patterns, and memory footprint. 

\subsection{Graph Ordering}
A number of graph ordering methods have been explored and demonstrated to be effective.

{\it Index-aware Ordering.}\quad It typically targets at the edge array layout. The basic idea is 
to sort the edges based on either the source vertex indices or \hl{the} destination vertex indices. 
Sorting the edges in an ascending manner generally improves 
the data locality because the neighboring vertex property can be pre-fetched and probably reused\upcite{13-GRAPHH}. 
In the graph processing, source vertex property will be read and destination vertex property will be 
updated accordingly. Therefore, reading overhead can be reduced if the edges are sorted by source vertices. 
Similarly, the writing process can be more efficient if the edges are sorted by the destination vertices\upcite{29-Zhou-Highthroughput}. As demonstrated in 
\cite{26-FPGP,30-ForeGraph,1-GRAPHICIONADO}, a hybrid \hl{index-aware} sorting method that balances both the source vertices and 
destination vertices \hl{can outperform the methods that only consider the source vertex or the destination vertex}. 

{\it Degree-aware Ordering.}\quad This method takes the vertex degree as the sorting metric. 
Sorting the vertices based on vertex degree in descending order brings
multiple benefits\upcite{36-DegreeBFS}. As high-degree vertices are more likely to be accessed, good data locality can be observed if 
high-degree vertices are placed nearby. In addition, it balances the workloads as well\upcite{38-FASTCF} when the 
graph is processed in parallel. The degree-aware ordering method 
applies to both baseline graph layouts\upcite{37-khoram-FPGAHMC}\hl{, i.e., the edge array and the compressed adjacency list}.

{\it Conflict-aware Ordering.}\quad This method is to reduce the data access conflict during parallel graph processing. 
ForeGraph\upcite{30-ForeGraph} proposes to interleave the edges such that memory level parallelism 
can be explored more efficiently. Different from the interleaving method, AccuGraph\upcite{40-YaoPACT} 
reorders the edges of the whole graph such that the destination vertices of the edges read in each 
cache line are distributed evenly over the on-chip memory banks. In this case, the parallel destination 
vertex updating has fewer conflicts.

{\bf Remarks.} Graph ordering methods focus on changing the order of the graph data organization. The reordered graph can be directly used by the graph 
accelerators without any modification. Nevertheless, the graph ordering usually requires global sorting 
and the pre-processing overhead, which can be costly. 

\subsection{Graph Partitioning}

Graph partition 
makes it possible to fit the graph into the limited on-chip memory of a graph accelerator.
The major graph partition strategies in graph accelerator designs can be roughly 
divided into four categories as shown in Table 1. 

\tabcolsep 12pt
\renewcommand\arraystretch{1.3}
\begin{center}
{\footnotesize{\bf Table 1.} Partitioning Schemes of Graph Accelerators}\\
\vspace{2mm}
\footnotesize{
\begin{tabular*}{\linewidth}{lp{4cm}p{4cm}}\hline\hline\hline
Partitioning Schemes & Graph Accelerators
\\\hline
Source-Oriented & {\cite{40-YaoPACT,29-Zhou-Highthroughput,9-RPBFS17,11-RPBFS18,15-WANG-MESSEGEBFS,22-UmurogluHybridBFS,34-ZhouCPUFPGA}} \\
Destination-Oriented &  {\cite{1-GRAPHICIONADO,26-FPGP,4-TUNAO,13-GRAPHH,12-GRAPHP}} \\
Grid & {\cite{30-ForeGraph,10-GRAPHR,39-HyVE}} \\
Heuristic & {\cite{2-EAA,6-TEMPLATEGAA,7-TESSERACT,38-FASTCF,37-khoram-FPGAHMC,5-GAA,25-GraphSoC}}
\\\hline\hline\hline
\end{tabular*}
\\\vspace{1mm}\parbox{8.3cm}{}
}
\end{center}

\vspace{1mm}

{\it Source-oriented Partition.}\quad 
The source-oriented partition methods typically have disjoint source vertices in each partition. 
All outgoing edges are associated with the partition's source vertices. The destination vertices will be included in 
the corresponding partition. Particularly, the source vertex indices in each partition are 
usually continuous to ensure sequential memory accesses. With the source-oriented 
partition, it is convenient to determine the partitions that need the updated vertex property in the 
graph processing. Nevertheless, different partitions may be in conflict with destination vertex update. To address this problem, \cite{29-Zhou-Highthroughput} 
proposes to synchronize through messages and resolve the data dependency through a specific computing unit.

{\it Destination-oriented Partition.}\quad 
The destination-oriented partition is similar to the source-oriented partition.
Basically the partitions have disjoint destination vertices. Therefore, each partition can be 
updated independently while reading the source vertex property for each partition is mostly 
random. Graphicionado\upcite{1-GRAPHICIONADO} adopts this partition method to ensure that 
each partition can be fitted to the small scratchpad memory. Low-latency high-bandwidth scratchpad memory 
can be fully utilized. GraphP\upcite{12-GRAPHP} also applies this partition. They aim at reducing the 
communication between the partitions on different accelerators such that the communication among the HMC cubes 
can be improved.

{\it Grid Partition.}\quad
The grid partition of graph in graph processing systems was first introduced in GridGraph\upcite{Gridgraph} which presented an efficient graph data layout and was widely absorbed into designs for graph processing accelerators\upcite{30-ForeGraph,10-GRAPHR}. Grid partition is essentially a \hl{two-dimensional} partition method, which can be considered an extension of the \hl{one-dimensional} partition, like source-oriented partition and 
destination-oriented partition\upcite{30-ForeGraph,10-GRAPHR}. First, it divides both the source vertices and \hl{the}
destination vertices into continuous segments. Then it forms a \hl{two-dimensional} array of cubes. Each cube includes the source vertex set, the 
destination vertex set, and all the edges whose source vertex and destination vertex belong 
to the source vertex set and destination set, respectively. The grid partition produces finer grained partitions.
The partitions have both sequential source vertices and destination vertices. 
ForeGraph\upcite{30-ForeGraph} uses this method to make best use of the limited on-chip memory of FPGAs. 
In particular, it optimizes the read order of partitions such that the partition loading and processing can be overlapped. 
This method is also used in GraphR\upcite{10-GRAPHR} and helps explore the ReRAM features for both high-performance low-power graph acceleration.  

{\it Heuristic Partition.}\quad
Unlike the above partition methods, many heuristic graph partition methods have been intensively explored, especially for 
conventional CPU-based graph processing systems. These partition methods follow various heuristic metrics \hl{to reduce 
the communication, to improve locality, or to provide better load balance}. Some of them are also applied for the graph 
accelerator design. For example, a \hl{hash-based} partition algorithm is used to achieve partitions with balanced 
vertices and edges in \cite{2-EAA}. A \hl{clustering-based} partition algorithm is adopted for better 
locality in \cite{37-khoram-FPGAHMC}. A multi-level partitioning algorithm is adopted in 
FASTCF\upcite{38-FASTCF} and is also demonstrated to be efficient for stochastic-gradient-descent-based 
collaborative filtering.

{\bf Remarks.}
Graph partition brings multiple benefits to graph accelerator design. In particular, 
it allows the graph accelerator to explore the small yet low-latency high-bandwidth on-chip memory. 

Graph preprocessing benefits the graph accelerator on many aspects including better data locality, more efficient memory access patterns, \hl{higher task-level parallelism}, and even fewer memory accesses. In general, it is a critical step to improve the performance of the graph processing accelerators, and even affects the accelerator design choices. While some preprocessing approaches are extremely time-consuming, it is still an open problem on how to achieve a better balance between the overhead and performance benefits in many practical scenarios as pointed out in \cite{Malicevic-Everything}. 

\section{Parallel Graph Computation}
The core component of a graph processing accelerator is how to handle the preprocessed graph data in Section~3 with massive parallellism. Considering intertwined data dependencies of graphs, this often requires non-trivial technical innovation, involving matched parallel iterative paradigms, dedicated hardware acceleration \hl{and} sophisticated co-codesigns. Fig.2 outlines the taxonomy of parallel graph computation.

\setcounter{figure}{1}
\begin{figure*}[!htb]
\centering
  \includegraphics[scale=0.9]{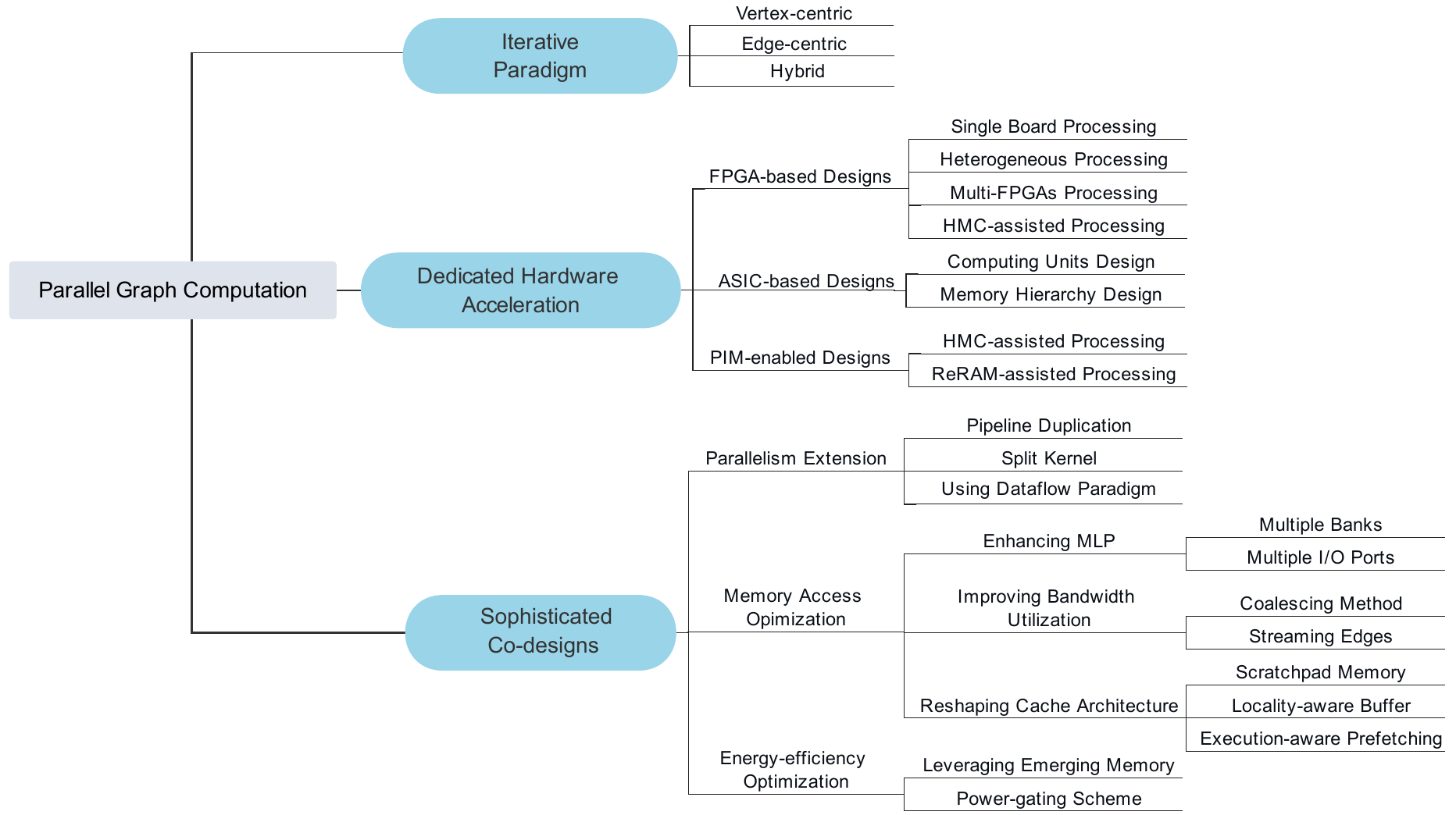}\\
  \caption{A taxonomy of parallel graph computation. 
  }
\end{figure*}
\baselineskip=18pt plus.2pt minus.2pt
\parskip=0pt plus.2pt minus0.2pt

\begin{itemize}[leftmargin=*]
\item  {\it Iterative Paradigm}. Iterative paradigm is used to express the process of how vertices and edges run. It defines the basic data access and computational pattern of graph program. Typical iterative paradigms in existing graph accelerators can be categorized into \hl{three approaches: the vertex-centric approach, the edge-centric approach, and the hybrid approach}. They decouple the associated dependencies within graphs as much as possible, and further explore the potential parallelism of graph processing.

\item \textit{Dedicated Hardware Acceleration}. Different kinds of dedicated hardware platforms can be used to accelerate graph analytics. Existing graph processing accelerators are basically built upon three types of hardware platforms: FPGA, ASIC, and PIM. These emerging architectures can be used to architect efficient memory hierarchy and computing units for higher performance and energy efficiency.

\item \textit{Sophisticated Co-designs}. Sophisticated co-designs usually combine the hardware and software optimizations to exploit the hardware potentials. They often focus on three aspects: parallelism extension, memory access optimization, and energy efficiency optimization. Most of these co-designs can be commonly used on \hl{different kinds of hardware} to achieve high performance and energy efficiency.
\end{itemize}

\subsection{Iterative Paradigm}
Graph has complex data dependencies between vertices. Designing efficient iterative paradigms \hl{is} important to decouple these associated dependencies as much as possible by exploring the common computational pattern surrounding vertices and/or edges. Existing iterative paradigms for graph processing can be basically divided into two subcategories: vertex-centric approach and the edge-centric approach. The vertex- and edge-centric approaches not only concern the expressiveness and abstraction of graph algorithms but also impact the design of graph data layout, preprocessing and computation. A few graph accelerators also have made a hybrid attempt for embracing the best worlds of both modes. Table 2 summarizes the related work with different iterative paradigms.

{\it Programming Model.} Programming model is used for effectively express the graph algorithms. It \hl{abstracts} the common operations in various graph algorithms and \hl{alleviates} the effort for programmers to write their applications. According to the iterative paradigms, there are vertex-centric programming model and edge-centric programming model. These two models can be combined as the hybrid model to take advantages of both paradigms. 

\tabcolsep 12pt
\renewcommand\arraystretch{1.3}
\begin{center}
{\footnotesize{\bf Table 2.} Iterative Paradigms of Graph Accelerators}\\
\vspace{2mm}
\footnotesize{
\begin{tabular*}{\linewidth}{lp{4cm}p{4cm}}\hline\hline\hline
Iterative Paradigm & Graph Accelerators
\\\hline
Vertex-Centric & {\cite{8-GRAPHPIM,40-YaoPACT,1-GRAPHICIONADO,20-CYGRAPH,26-FPGP,2-EAA,4-TUNAO,6-TEMPLATEGAA,7-TESSERACT,32-FPGAHMCBFS,36-DegreeBFS,37-khoram-FPGAHMC,15-WANG-MESSEGEBFS,12-GRAPHP,5-GAA,16-GRAPHLET,17-BetkaouiBFS,18-BetkaouiAPSP,19-GRAPHGEN,21-AttiaAPSP,25-GraphSoC,27-GraVF,33-FPGADATAFLOW}} \\
Edge-Centric &  {\cite{29-Zhou-Highthroughput,30-ForeGraph,10-GRAPHR,28-GraphOps,13-GRAPHH,38-FASTCF,39-HyVE,23-ZhouSSSP,24-ZhouPAGERANK,41-GRAFBOOST}} \\
Hybrid & {\cite{34-ZhouCPUFPGA}} 
\\\hline\hline\hline
\end{tabular*}
\\\vspace{1mm}\parbox{8.3cm}{}
}
\end{center}
\vspace{1mm}

\begin{itemize}[leftmargin=*]

\item {\it Vertex-centric Programming Model.} Graph algorithms expressed with this model handle the graphs by following ``Think like a vertex" philosophy\upcite{Pregel}. It describes a graph program for each vertex, including computational operations and data transmission between their neighbors via edges. Since each vertex is processed independently, parallelism can be therefore guaranteed by simultaneously scheduling these vertices without data dependencies.

\item {\it Edge-centric Programming Model.}\quad
X-Stream\upcite{X-Stream} is the first work to use edge-centric programming model to handle graph edges. Unlike the vertex-centric model, this model describes a graph program for each edge. An edge is processed with three steps: 1) collect the information of its source vertices, 2) update its value, and 3) send this value to its destination vertices. It is clear that this model removes the random accesses to edges via sequential streaming of each edge to the chips. 

\item {\it Hybrid Programming Model.}\quad
Alternative is to use a hybrid method by switching between vertex- and edge-centric programming models for the purpose of taking advantages of both models\upcite{34-ZhouCPUFPGA}. The vertex-centric model is responsible for the situation when \hl{the} active vertex ratio is relatively high. In contrast, the edge-centric model is intended to cope with the case that active vertex ratio is relatively low. Clearly, model switching decision can be made according to the active vertex ratio (among all vertices). The threshold can be decided by the ratio of bandwidth. 

\end{itemize}

{\it Data Layout Selection.} Systems implemented in vertex-centric approach typically iterate over the active vertices and execute the vertex program on them at each iteration. For each given vertex, its neighbours are visited to complete the computation. This kind of implementation usually requires a fast scan for edges of given vertices. As a consequence, as presented in Section 3.1, the compressed adjacency lists (CSR/CSC) are suitable for vertex-centric model because the assoiated edges of a vertex can be found easily\upcite{GraphChi, 2-EAA}.

Similarly for the edge-centric approach, which iterates over all the edges to implement the edge program for each of the edge, a fast sequential scan of edges is demanded. To process an edge, information of the end vertices also \hl{needs} to be indexed directly. Therefore, the edge array presented in Section 3.1 is intuitively fit for systems in edge-centric approach\upcite{X-Stream,29-Zhou-Highthroughput}.

{\it Preprocessing Considerations.} Initially, the graph data is usually stored in the disk as edge files where the edge is represented as a pair of corresponding source and destination \hl{vertices}. During preprocessing phase, edge files are converted into the appropriate data layout according to programming models. As discussed in Section 3, preprocessing involves graph partitioning, reorganization and ordering. The complexity of preprocessing also varies for different data layouts.

For vertex-centric approach, the edge file is converted into the format of adjacency lists. Typically, the edges are sorted by source or destination vertex following by index creation that maintains the edge position in the edge array for each vertex. As for edge-centric approach, the edge array is usually loaded directly without specialized data formatting and conversion\upcite{X-Stream,29-Zhou-Highthroughput}. A detailed research about the cost on preprocessing is presented in \cite{Malicevic-Everything}. Generally, the preprocessing cost on vertex-centric approach is higher than edge-centric one.

{\it Computation Overhead.} Vertex- and edge-centric approaches have different computation patterns as dissucssed before. In vertex-centric approach, the computation is executed for each vertex which iterates over the neighbors of a given vertex. In edge-centric approach, the edges are executed as a stream. At this point, the workload characteristics and cache (miss-rate) metrics are significantly different for two approaches\upcite{Malicevic-Everything}.

For workload analysis, vertex-centric approach supports selective scheduling on only a subset of vertices for each iteration while edge-centric approach requires a scan of the whole edges, which means that the edge-centric approach induces more computations than vertex-centric approach.

Cache behaviours are also different between these two approaches. In vertex-centric approach, the processed vertices can be (locally) cached while it introduces more random accesses by traversing the frontier. In edge-centric approach, edges can be prefetched for better use of cache, but it causes more random accesses to vertices. Their actual performance may be significantly different, and largely \hl{depends} on the inherent topology of the graph and features of graph algorithms.

\setcounter{table}{2}
\tabcolsep 9pt
\begin{table*}[!htb]
\centering
\caption{\label{ti} Overview of Different Iterative Paradigms}\vspace{-2mm}
{\footnotesize
\begin{tabular*}{\linewidth}{c p{3cm}p{2.3cm}p{3.2cm}p{3.2cm}}
\hline\hline
Iterative Paradigm & Programming Model & Data Layout & Preprocessing & Computation Overhead\\
\hline
Vertex-centric & Iterate over vertices & CSR/CSC & Partitioning; Ordering; Reindexing; Higher cost &  frontier-based; random accesses to edges\\ \hline
Edge-centric   & Iterate over edges & Edge array/COO & Partitioning; Ordering; Lower cost & All edges need to be scanned; random accesses to vertices \\ \hline
Hybrid       & A mix of vertex- and edge-centric model & Mixed data structures & Sophisticated preprocessing & Model switch\\
\hline\hline
\end{tabular*}
}
\end{table*}

Generally, the vertex-centric approach introduces more random accesses to edges while the edge-centric approach causes more random accesses to vertices. To improve the cache behaviours, optimizations can be applied to these two models, e.g., organizing edge arrays into grids can improve the cache locality\upcite{Gridgraph}.

\textbf{Discussions.} Table 3 compares different paradigms from multiple aspects. It is difficult to judge which approach is better because the performance is usually not the same case when different kinds of graph applications are introduced. The authors in \cite{Malicevic-Everything} make a comprehensive comparison of these two approaches when different approaches and graph algorithms are included.

Vertex-centric paradigm has been widely used to drive many graph accelerators\upcite{1-GRAPHICIONADO,26-FPGP,2-EAA,19-GRAPHGEN}, because of its expressive potentials to easily represent various \hl{kinds} of graph algorithms, and the high parallelism in the grain of vertex. However, in the vertex-centric paradigm, there can be random accesses to edges, resulting in potentially heavy memory access overhead.

Edge-centric paradigm is usually used by existing graph accelerators for improving the utilization of their limited memory bandwidth\upcite{29-Zhou-Highthroughput,30-ForeGraph, 38-FASTCF}. However, the point is that edge-centric paradigm is lack of flexible scheduling potential in contrast to the vertex-centric one. Almost all of edges have to be processed multiple times to complete the whole process. In addition, this paradigm may also lead to a large number of random accesses to vertices. Thus, additional optimizations are often cooperatively needed, e.g., fined grained partitioning and tailored vertex update strategies\upcite{30-ForeGraph,10-GRAPHR}.

For graph processing accelerators, the selection and design of iterative paradigm for graph processing accelerator must also ensure that: 1) programming is easy to use and understand for graph algorithm representation; 2) parallelism is easy to expose and exploit for high throughput and fast hardware development. It is also important to dedicate the accelerators according to the features of applications. Note that advantages can be combined by incorporating hybrid approaches into a design for better performance.

\subsection{Dedicated Hardware Acceleration}
Existing graph processing accelerators can be built upon various kinds of hardware platforms. Typical hardware accelerators usually adopt only the traditional customized hardware platforms, i.e., FPGAs and ASICs, and have made few modifications on existing \hl{computing} logic and memory architectures (e.g., DRAM). Some accelerators have re-built their architectures with in-situ computation without excessive data movement, e.g., HMC and ReRAM devices, which is also known as PIM-enabled accelerators. Different hardware configurations have different considerations for performance acceleration. We next review technical advances of these state-of-the-art graph processing accelerators.

\subsubsection{FPGA-based Designs}
FPGA is an integrated circuit that enables designers to repeatedly configure digital logic in the fields after manufacturing, also called \hl{field-programmable}. The configuration of FPGAs is generally specified via low-level hardware description languages, e.g., Verilog\upcite{VERILOG} and VHDL\upcite{VHDL}. FPGAs are mostly adopted in graph processing accelerators.

{\it Internal Characteristics of FPGAs.}\quad
There are different kinds of programmable resources on FPGAs, e.g., programmable \hl{Logic Element (LE), Static Random Access Memory (SRAM), Flash and Block RAM (BRAM)}. However, the fact is that these resources are usually limited to a small number.
FPGA can offer high parallelism by architecting these resources with a pipelined \hl{Multiple Instructions Single Data (MISD)} model. 
Multiple data can be processed simultaneously at different pipeline stages. Multiple pipelines can be easily duplicated for parallel processing. 

{\it Existing Efforts on FPGAs.}\quad
A graph program is usually designed into a circuit kernel as \hl{the} basic processing unit according to the programming model (as discussed in Section 4.1), which defines the execution pattern\upcite{38-FASTCF, 18-BetkaouiAPSP}. These kernels can be easily reconfigured on FPGAs for different algorithms. For building the efficient memory subsystem, a wide spectrum of previous studies make non-trival efforts for the efficient bandwidth utilization of on-chip BRAMs and the off-chip \hl{memories}. BRAMs \hl{provide} high bandwidth and low memory latency for randomly accessed vertices. For improving the locality of vertices on the BRAM, fine grained partitioning and dedicated data placement \hl{strategies} are needed to increase the reuse rate of vertices on the BRAM\upcite{26-FPGP, 30-ForeGraph, 36-DegreeBFS}. As for improving the utilization of off-chip bandwidth, edges can be streamed sequentially from the memory\upcite{29-Zhou-Highthroughput}.

A number of studies extend to integrate multiple FPGAs into a cluster so as to support large graphs\upcite{20-CYGRAPH, 32-FPGAHMCBFS}. FPGAs with integrated soft-cores are also presented, which can process the graphs in a distributed manner on a single FPGA board with high parallelism\upcite{25-GraphSoC}. Heterogeneous architectures \hl{are} also adopted where the FPGA and the CPU are connected through cache-coherent interconnect. FPGA can access the host memory without the interruption of CPU. These two processors can easily cooperate with each other to process large graphs with higher parallelism than a single FPGA board\upcite{34-ZhouCPUFPGA}. 

There \hl{are} also a number of studies that aim at exploring the out-of-memory execution on FPGAs for large graphs. The data can be directly streamed from the disks or flashes to the processing units on the FPGA board in these scenarios\upcite{26-FPGP, 30-ForeGraph}. Recently, \hl{Near-Data Processing (NDP)} \hl{is} cooperatively used to enhance the power of FPGAs for graph processing by off-loading workloads to the integrated HMCs. This provides significantly high bandwidth and parallelism \upcite{32-FPGAHMCBFS,37-khoram-FPGAHMC, 35-ExtraV}.

\subsubsection{ASIC-based Designs}
ASIC is an integrated circuit composed of electrical components, e.g., resistors. It is usually fabricated on a wafer composed of silicon or other semiconductor \hl{materials} that \hl{are} customized for a particular use. ASICs are very compact, fast, and low power. Compared to FPGAs, their functions are hard-wired at the time of manufacture. It is not allowed to change the functionality of a small part of the circuit. 

{\it ASIC Designs for Graph Analytics.}\quad 
Due to the fixed circuit limitation, ASIC-based graph processing usually utilizes the expressive Gather-Apply-Scatter (GAS) model\upcite{PowerGraph} to form the circuit\upcite{2-EAA,4-TUNAO}. Each phase is implemented as a hardware module, and runs in parallel with wires that \hl{connect} different modules. In order to support various graph algorithms, a reconfigurable block can be integrated for users to define the update functions for flexibility. 

As for the memory hierarchy, these graph accelerators commonly adopt the scratchpad memory to replace traditional cache. The scratchpad memory acts as a content addressable cache and can be controlled manually. Graphicionado\upcite{1-GRAPHICIONADO} uses the eDRAM as the scratchpad memory to store graph data that needs frequent random accesses, e.g., the destination vertices. Dedicated caches of different kinds of graph data are designed in \cite{2-EAA} according to the access features. Since these memory resources can be tightly connected to the processing units in an efficient way, ASIC-based graph accelerators can achieve high throughput on \hl{the} chip. 

\subsubsection{PIM-enabled Designs}
Different from traditional hardware designs, research on PIM-enabled architectures usually \hl{adopts} emerging paradigms for acheiving impressive performance results by integrating processing units into the memory. It can provide the extremely high bandwidth and low memory access latency with energy saving. The PIM-enabled acceleration is often implemented by leveraging emerging memory devices, e.g., HMC and ReRAM, both of which integrate the in-situ computation in the memory.

{\it HMC-assisted Graph Processing.}\quad
The HMC has multiple DRAM dies stacked on top of a logic layer that can provide the ability of computation with high memory access parallelism and sufficient instructions for supporting graph processing. As in Fig.3, the DRAM dies are connected via \hl{the} Through-Silicon-Via (TSV). Storage space in HMC is organized as vaults. \hl{The vault} is a vertically connected stack of multiple partitions from different DRAM layers. The logic layer is also distributed to different vaults. With multiple DRAM channels for each vault, HMC can provide significantly high memory-level parallelism. HMC can also be easily scaled to consist of a cluster topology of HMCs\upcite{HMC-connect}.

\begin{center}
\includegraphics[width = 8.3cm]{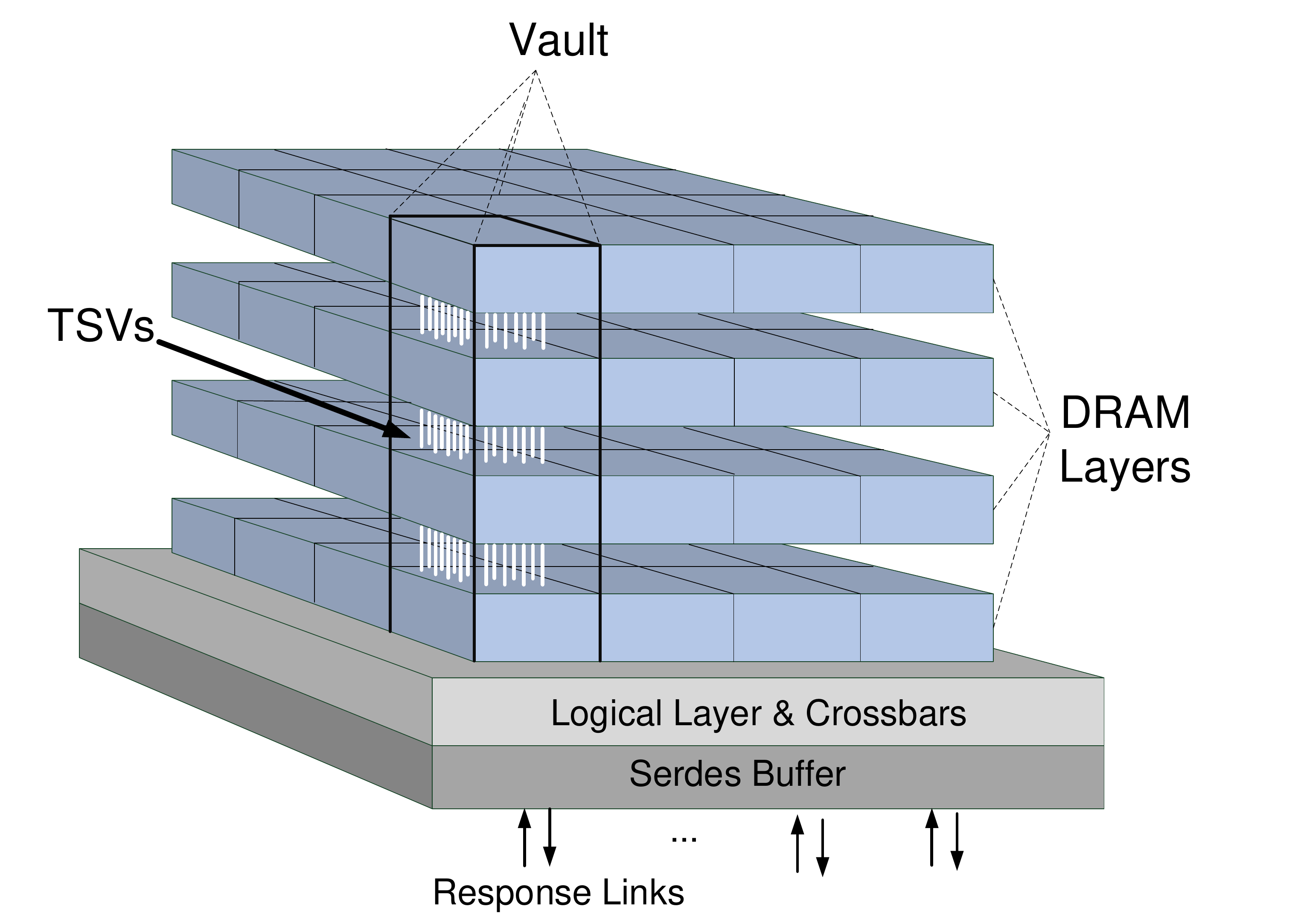}\\
\vspace{2mm}
\parbox[c]{8.3cm}{\footnotesize{Fig.3.~}  An illustrative example of HMC architecture}
\end{center}

The logic layer of each vault can work as a soft-core with sufficient instruction sets. For better supporting graph processing, instructions have to be re-constructed. Tesseract\upcite{7-TESSERACT} integrates common instructions of graph algorithms and achieves high performance through multiple HMCs. GraphPIM\upcite{8-GRAPHPIM} deigns specialized atomic instructions in HMCs. Besides, graphs are processed in a distributed manner between HMCs. Vertex-cut partitioning is also used to reduce the communication cost between HMCs\upcite{13-GRAPHH, 12-GRAPHP}.

{\it ReRAM-assisted Graph Processing.} ReRAM is a kind of non-volatile RAM with the enabled \hl{computing} ability by changing the resistance across a dielectric solid-state material\upcite{ReRAM}. A ReRAM cell is with high density, low read latency and high energy efficiency\upcite{ReRAM2}. The ReRAM cells can be connected as a dense crossbar architecture to provide high parallelism and memory capacity. Generally, the graph can be represented as a matrix which can be naturally mapped to ReRAM cells. Each cell stores an edge or a vertex. When input voltages are applied to certain rows of the cells array, the stored values of each row will multiply the relevant input value. The stored values of each columns will be then added together. These features make it possible to realize efficient graph processing on ReRAM.

The potential of ReRAM for efficient computation and storage is still under-studied significantly. To our best knowledge, GraphR\upcite{10-GRAPHR} is the first work to use ReRAM to speedup the graph computation. It transfers \hl{the} vertex program or \hl{the} edge program in graph processing to a Sparse Matrix-Vector Multiplication (SpMV) format. However, graph algorithms need to be manually justified for mapping the computational pattern of ReRAM. It is worth noting that there is also a tradeoff between the utilization and throughput due to the limited ReRAM cell size. 
An ideal situation is that every entity within a matrix is useful for computation for high parallelism. Nevertheless, due to the sparsity of graph data, in a ReRAM block there may be only a few useful edges that are non-zero, causing the fact that a large number of ReRAM cells are underutilized. Extra efforts are still needed to balance this tradeoff.

{\bf Summary.}
Considering the reconfigurable feature, FPGA-based designs can handle various kind of graph algorithms flexibly. FPGA can also provide sufficient interfaces to process large graphs for scale-out efficiency. Massive parallelism can be easily achieved when these resources are in good use. Unfortunately, the resources on FPGAs are limited for existing commodity FPGA boards. The frequency is also relatively low to maintain correctness of execution. These may influence the performance of graph processing.

ASIC designs can provide efficient hardware organizations without the limitation on types and numbers of hardware resources. ASICs can be designed in a relatively efficient way. For example, dedicated and accurate resources placement in ASCIs can be achieved while FPGAs usually have redundant and wasted resources on board. Besides, the ASIC can achieve a high frequency than FPGAs. High performance can be easily attained. However, once made, the ASIC chip is unable to be modified. It is usually difficult for ASICs to handle various graph problems. It is also difficult for ASICs to scale out.

PIM-based accelerators can scale well in both of the bandwidth and memory capacity. This feature can benefit the graph processing when large graphs are handled. The emerging memories adopted in \hl{PIM-based} accelerators usually have lower energy consumption than traditional DRAM. To handle generic graph analytics, the HMC provides the computing ability by special instruction sets \hl{executed} in the logic layer. ReRAM processes the graphs in the SpMV format with corssbars. These supports usually need many manual efforts to realize. There is still a lot of research space for \hl{PIM-based} graph accelerators. For example, the bandwidth can be underutilized due to the communication overhead in HMCs.

\subsection{Large-Scale Graph Processing Acceleration}

Real-world graph data size can easily exceed the on-chip/board memory capacity of graph processing accelerators. Most of existing accelerators only consider the case that the whole graph \hl{fits} into the no-chip/board memory. However, how to deal with larger graphs on accelerators is a vital issue for practical applications. There is an amount of work that \hl{has} taken this issue into account\hl{, and a series of solutions are further proposed}\upcite{20-CYGRAPH, 26-FPGP, 30-ForeGraph, 7-TESSERACT, 34-ZhouCPUFPGA, 41-GRAFBOOST}. These solutions can be typically divided into three categories\hl{: the out-of-core solution, the multi-accelerators solution, and the heterogeneous solution.}

{\it 1) Out-of-Core Solution.}\quad 

\hl{Unlike traditional CPU architectures that involve large main memory and often develop the out-of-core solutions based on the disks, graph accelerators typically have relatively small on-chip/board memory capacity. 
Therefore, toward graph accelerators, using any external storages or memories that can store large real-world graphs can be considered potentially-useful as their out-of-core solutions. 
Graph accelerators can use disks, flashes or other external storage devices to store extremely large scale graphs\upcite{GraphChi, X-Stream, Gridgraph, 26-FPGP, 41-GRAFBOOST}}. However, one of the most important issues for utilizing these devices is to reduce the transmission cost of I/Os between the disk and the DRAM since the \hl{bandwidths} of these devices is often significantly lower than DRAM. Streamlined processing schemes\upcite{X-Stream, 41-GRAFBOOST} and sophisticated partitioning methods\upcite{Gridgraph, 26-FPGP} can be explored to effectively reduce the overhead of memory accesses to these external devices. \hl{Recently, utilizing embedded processors or accelerators in SSDs has been proved to be another promising way to alleviate the overhead of data transmission and conversion\upcite{flash1, flash2, flash3}.}

\hl{Compared to disk-based solutions, utilizing large host memory enables graph accelerators to process large-scale graphs with better bandwidth-efficiency\upcite{17-BetkaouiBFS, 29-Zhou-Highthroughput, 35-ExtraV}.
Emerging computing platforms offer the great opportunity for graph accelerators to access the main memory conveniently via specialized interconnections\upcite{emerging-platforms}.
However, it is also vital to optimize the I/Os between graph accelerator and main memory, since long memory latency for data movement often dominates the overall efficiency due to slow I/O interfaces and extra efforts on memory management\upcite{pointer-chasing-fpga}.
Existing memory subsystems and their memory access parallelism are strongly in need of technological \hl{innovation}. }

There also have emerged some studies regarding graph processing accelerator designs for large-scale graph processing. FPGP\upcite{26-FPGP} incorporates the disks to extend the storage of FPGA and designs a streamlined vertex-centric graph processing framework to improve the utilization of the sequential bandwidth of disks. A dedicated on-chip cache mechanism is used to reduce the accesses to disks. Then the large graph is specially partitioned in order to fit for the processing scheme. GraFBoost\upcite{41-GRAFBOOST} adopts the flash to scale to much larger graphs and mainly \hl{focuses} on optimizing the random accesses. The key component is a sort-reduce module that converts small random accesses into large block sequential accesses to the flash storage. \hl{It is mentioned that GraFBoost\upcite{41-GRAFBOOST} embeds the accelerator into the flash for better scalability. Similar methods have been explored to accelerate the processing in database\upcite{flash4, flash5}. ExtraV\upcite{35-ExtraV} further incorporates the main memory to improve the graph processing with SSDs. Note that host processors can be used together with its self-contained main memory in a heterogeneous solution to enhance the power of graph accelerators.}

{\it 2) Multiple Accelerators Extension.}\quad 

The whole graph needs to be partitioned to distribute different on-chip/board \hl{memories} of each graph processing accelerator. By considering the prohibitive communication overhead between graph accelerators, the multi-accelerator solution often needs the high-bandwidth connection between graph accelerators. The most important issue for this design is how to achieve a cost-efficient communication mechanism, and avoid data conflicts between graph accelerators. As a consequence, the appropriate graph partition methods are required and \hl{are} important to reduce the communication overhead\upcite{30-ForeGraph, 7-TESSERACT, 12-GRAPHP}. The inter-network design of graph accelerators is also vital to support the efficient cooperative computing\upcite{20-CYGRAPH, 13-GRAPHH}.

CyGraph\upcite{20-CYGRAPH} \hl{runs} BFS under a high performance reconfigurable computing platform, Convey HC-2, which \hl{constructs} a platform with FPGAs connected through a full crossbar to multiple on-board memories. These memories are connected as a shared memory that \hl{provides} large capacity and high total bandwidth. CyGraph optimizes the CSR representation to reduce the shared memory accesses and connects the FPGAs using a ring network to minimize the conflicts. ForeGraph\upcite{30-ForeGraph} instead \hl{uses} separated memories for each FPGA. Thus \hl{it avoids} the memory access conflicts among accelerators. These FPGAs are connected via dedicated inter-connections. Grid-like partitioning\upcite{Gridgraph} and dedicated on-chip data replacement schemes are adopted to achieve better locality for each FPGA board and thus reduce the communications.

As discussed in Section 4.2, emerging devices like HMCs not only provide the capability of processing in memory but also scale well. \hl{The cost} on communications among different HMC cubes dominates the performance\upcite{7-TESSERACT, 12-GRAPHP, 13-GRAPHH}. GraphP\upcite{12-GRAPHP} utilizes a source-cut partitioning method to significantly reduce the communication overhead. Generally, multi-accelerator solutions is similar to distributed processing under traditional platforms such that many optimizations on distributed graph processing can be applied to accelerators. Meanwhile, the features of different architectures need to be considered to provide the best scenario.

{\it 3) Heterogeneous Acceleration.}\quad 

As the rapid development of memory integration \hl{technologies} (e.g., 3D stacking), the host memory becomes large or even huge with trillions of capacity\upcite{Ligra, Galois}. As a consequence, leveraging the host-side memory is alternative to support large-scale graph processing. An intuitive and important question is how graph processing accelerator can interact with the host machine conveniently and efficiently. At present, efficient heterogeneous \hl{solutions} are still open \hl{questions}. A few studies propose to use the coherent memory interconnect technology to accelerate graph workloads with CPU and FPGA\upcite{34-ZhouCPUFPGA}. For supporting efficient cooperation, the dedicated memory subsystem is needed to alleviate the transmission overhead between the host and the graph accelerator. As a result, the data organization of graphs is the key to reduce the communication overhead. In order to avoid conflicts of computing devices, runtime scheduling schemes are also important for efficient task scheduling.

The authors in \cite{34-ZhouCPUFPGA} propose to accelerate graph processing under a heterogeneous architecture with CPU and FPGA. Hybrid vertex- and edge-centric models \hl{are} adopted in \cite{34-ZhouCPUFPGA} as discussed in Section 4.1 to fully utilize the processing power of CPUs and FPGAs. Generally, CPU is better for fast sequential processing while FPGA can be used to explore massive parallelism. Hybrid model can flexibly assign workloads to these two devices according to the parallelism of vertices in each iteration. In order to support this scheme, an optimized graph data structure is designed. As for memory coherency, dedicated on-chip memory buffers are designed on FPGA and the accesses to the host memory is controlled by a master thread on the CPU. Despite that the heterogeneous solution can extend the power of accelerators, the overhead to maintain the memory coherency might limit the performance. There is still a lot of research space for heterogeneous solutions. 

\subsection{Sophisticated Co-designs}
Graph processing accelerators often require a series of optimizations for fully exploiting their hardware potentials.
There also emerge a few co-optimization techniques at these aspects for high parallelism, lower memory access overhead, and better energy efficiency.

\subsubsection{Parallelism Extension}
The processing units in either ASIC- or FPGA-based graph processing accelerators are often organized in the form of pipelines. The instructions of graph algorithms are pipelined to offer high parallelism. PIM-based graph accelerators integrate the processing units inside the memory. Their efficiency can be scaled by simply enlarging the memory capacity. For better scalability, three optimization solutions below can be considered useful potentially.

{\it Pipeline Duplication.}\quad
An intuitive method to increase the throughput is to duplicate multiple pipelines for the parallel processing on more vertices and edges. This simple method has been widely used in a wide spectrum of previous work\upcite{1-GRAPHICIONADO,29-Zhou-Highthroughput,2-EAA,4-TUNAO,16-GRAPHLET,23-ZhouSSSP,3-NOVELSPMV}. Nevertheless, there still remain some potential problems that might prevent the scalable efficiency of multi-pipeline from expectation, which is significantly under-studied. For instance, 
considerable communication between pipelines may lead to the additional overhead via crossbars and controllers\upcite{1-GRAPHICIONADO,2-EAA}. In addition, there also exists workload balance issue that needs specialized data partitioning\upcite{1-GRAPHICIONADO,30-ForeGraph}.

{\it Split Kernel.}\quad
Alternative is to split a big, whole processing stream into many small kernels that can be then considered being executed in parallel. This is often done by decoupling the modules of data access and computation, and then making them executed in parallel. In this way the data access module is responsible for accessing graph data. The computation module uses the data to conduct user-defined computations. 
For example, by using GAS model, \cite{20-CYGRAPH,2-EAA,4-TUNAO} create specialized execution circuits. Each module is thus enabled to process a large number of vertices and edges concurrently. The SpMV-based accelerator\upcite{3-NOVELSPMV} also decouples the matrix access from the computation. This method explores the task-level parallelism but extra scheduling mechanism are needed to ensure the correctness.

{\it Using Dataflow Paradigm.}\quad
Vertex dependencies of graph can stall the pipelines and decrease the instruction level parallelism. How to reduce the impact arising from data dependencies remains an open problem for increasing the number of Instructions per Cycle (IPC). 
One viable solution for solving this problem is to leverage the dataflow paradigm\upcite{28-GraphOps,33-FPGADATAFLOW, SCI-DATAFLOW}, which forms a directed graph of the operations according to the data dependency between two adjacent operations. The output dependency and control dependency in graph processing can be then significantly eliminated\upcite{33-FPGADATAFLOW}. GraphOps\upcite{28-GraphOps} uses dataflow model to form the data path of different processing blocks. Their overhead of \hl{controlling feedback} can be therefore alleviated.

\subsubsection{Memory Access Optimization}
For graph processing, memory accesses often dominate overall execution time. Designing \hl{an} efficient memory subsystem is crucial for the graph processing accelerator, particularly for memory access efficiency\upcite{1-GRAPHICIONADO, 2-EAA}.

{\it 1) Enhancing Memory-Level Parallelism (MLP).}

The MLP can be represented as the number of outstanding memory requests supported at the same time. Higher MLP can reduce the total memory access time for data-intensive applications as graph processing. It usually needs the memory devices to support enough concurrent memory requests. There are two ways \hl{to enhance} the MLP.

\textit{Multiple Banks.} One method to increase the MLP is using multiple banks. DRAM is composed of many independent banks. Utilizing the parallelism of these banks can significantly improve the memory level parallelism\upcite{16-GRAPHLET,17-BetkaouiBFS,18-BetkaouiAPSP}. The memory banks are connected to the processing units directly through crossbars. They can be accessed concurrently.

\textit{Multiple I/O Ports.} Another method is to design multiple I/O ports for a memory block\upcite{29-Zhou-Highthroughput,19-GRAPHGEN,23-ZhouSSSP}. By increasing the I/O ports, multiple memory requests can run concurrently. Usually the number of ports can be manually organized on \hl{the} scratchpad memory. High MLP can be attained when the number of ports are equal to the number of processing units\upcite{1-GRAPHICIONADO}. BRAMs on FPGAs can also be manually controlled to achieve this goal\upcite{29-Zhou-Highthroughput}. These BRAMs are usually combined together to form a memory block with multiple I/O ports.

{\it 2) Improving Bandwidth Utilization.}

The memory bandwidth utilization here means the valid values ratio per transfer. Random accesses in graph processing usually \hl{cause} low ratio of valid values and result in much wasted bandwidth. Improving the memory bandwidth utilization can reduce the total number of memory accesses. There are mainly two effective methods for improving the bandwidth utilization.

\textit{Coalescing Method.} Coalescing means combining multiple transfers of small items into fewer large ones. This method is widely adopted in graph accelerators\upcite{29-Zhou-Highthroughput,32-FPGAHMCBFS,19-GRAPHGEN,23-ZhouSSSP,24-ZhouPAGERANK}. For example, if the memory requests are adjacent in a vertex or edge list\hl{,} these requests can be \hl{coalesced} as one request for a block. Otherwise there may exist several random accesses that lead to the wasting of bandwidth\upcite{19-GRAPHGEN}.

\textit{Streaming Edges.} Streaming edges means that the edges are sequentially accessed from the memory to accelerator\upcite{29-Zhou-Highthroughput}. Random accesses of edges can be reduced. In a vertex-centric model, the edges of a vertex can be streamed to the chip\upcite{1-GRAPHICIONADO}. This method can fully utilize the bandwidth in the edge-centric model. However, the edges may need to be reordered so as to run in a more efficient fashion\upcite{29-Zhou-Highthroughput,30-ForeGraph}.

{\it 3) Reshaping Cache Hierarchy.}

Poor locality of graph processing makes the current cache hierarchy lack of efficiency. High cache miss rate will increase the memory access latency, which would cause the under-utilization of computing resources. Reshaping the cache hierarchy means designing new cache architectures and mechanisms for graph processing features.

\textit{Scratchpad Memory.} Scratchpad memory is used as an addressable cache that can be explicitly controlled. The scratchpad memory is closed to the graph engines. It can provide high performance for data access\upcite{13-GRAPHH,FPGACAM, SCI-SPM}. Graphicionado\upcite{1-GRAPHICIONADO} uses scratchpad memory to store the temporary vertex property array and edge offset to optimize the random data accesses. Similarly, \cite{2-EAA} also designs different kind of caches for vertices, edges, and other graph information according to their access behaviors. 

\textit{Locality-aware Buffer.} Locality-aware buffer is \hl{a kind of specialized cache} for graph data with relatively good locality, e.g., the high degree vertices. High degree vertices in a power-law graph have \hl{higher probability} to be accessed many times. These vertices can be cached to improve performance\upcite{4-TUNAO}. FPGP\upcite{26-FPGP} and ForeGraph\upcite{30-ForeGraph} improve the locality of vertices using grid-like partitioning methods, and design special on-chip buffers for vertex subsets, which can be thus accessed fast in reuse.

\textit{Execution-aware Prefetching.} This method prefetches the graph data according to the execution requirements. It avoids the inefficiency of fixed traditional cache prefetching mechanism.  For example, in vertex-centric model, the source vertex list and its corresponding edge list can be prefetched sequentially\upcite{7-TESSERACT}. The key is to exploit the access patterns of different kinds of graph data during the execution, and further design appropriate prefeching mechanism to reduce the memory latency.

\subsubsection{Energy efficiency Optimization}
The performance of graph accelerators can be measured as Traversed Edges per Second (TEPS). Energy efficiency can be further defined as TEPS per Watt (TEPS/W). Existing graph processing accelerators can provide significantly high performance by dedicated circuits with inherent low-energy consumption. However, most of graph programs have a high memory-access-to-computation ratio. For example, the energy results show that PageRank consumes over 60\% energy on memory\upcite{Gao-PIMenergy}. Optimizations on memory consumption can further improve the energy efficiency. Nowadays, there are two simple yet effective ways to improve the memory energy consumption. 

{\it Leveraging Emerging Memory Technologies.} A number of emerging memory technologies integrate the computing logic inside the memory, e.g., HMC\upcite{8-GRAPHPIM,7-TESSERACT,13-GRAPHH,12-GRAPHP} and ReRAM\upcite{10-GRAPHR,39-HyVE} as described previously. This architectural reformation can conduct the in-situ computation along side the data. It naturally avoids the frequent data movement for energy saving. 
At this point, we can easily replace traditional DRAM by leveraging these emerging memory devices. 

{\it Power-gating Schemes.} Power-gating is a widely used technology that powers off the idle logic circuits to save the energy. This scheme is suitable for memories that can be manually controlled\upcite{29-Zhou-Highthroughput, 39-HyVE}. For example, it can be applied to BRAMs on FPGAs, which are the key for improving the overall FPGA energy consumption in graph processing accelerators\upcite{29-Zhou-Highthroughput}. The BRAM is selectively activated and deactivated via the enabled ports. A BRAM module is only activated when the required data is stored. When the edges of a vertex are stored in the same BRAM module, BRAM only needs to be activated once to traverse these edges\upcite{29-Zhou-Highthroughput}. Similar \hl{strategies} can be used for ReRAM\upcite{39-HyVE} to save the energy for edge access by controlling the activation of ReRAM banks in a flexible way.

\section{Runtime Scheduling}

As discussed in Section 4.2, customized hardware circuits for graph processing generally involve specialized designs. This often enforces to design the tailored runtime scheduling to appropriately assign workloads and coordinate the processing units for providing the correct and efficient execution.
Unlike existing runtime schedulers on traditional processors, the runtime scheduling for graph accelerators may be necssarily needed to be implemented in the form of hardware circuits. This process usually needs to be transparent to users. Runtime scheduling usually involve three aspects of core components: the communication models, the execution modes, and the scheduling schemes.

\begin{itemize}[leftmargin=*]

\item \textit{Communication Model}. Communications commonly exist in graph processing accelerators among processing units. Communication models provides efficient ways for 
these processing units to communicate and cooperate with each other. Graph accelerators usually adopt two kinds of communication models: the message-based pattern and the shared memory pattern. These models present different features and can benefit from the optimization of information flows. 

\item \textit{Execution Mode}. The execution mode determines the scheduling order of operations. There are two kinds of execution modes that have been widely used for existing graph processing accelerators: synchronous execution and asynchronous execution. 

\item \textit{Scheduling Schemes}. Scheduling schemes defines the granularity and processing order of graph data. Existing work adopts three kinds of scheduling schemes: block-based scheduling, frontier-based scheduling, and priority-based scheduling. Flexibly using these scheduling schemes can help reduce the conflicts and improve the utilization of hardware resources. 

\end{itemize}

\subsection{Runtime Considerations}

For preserving the correctness and efficiency, runtime scheduling for graph processing accelerator needs to consider the following two major aspects.

\begin{itemize}[leftmargin=*]
\item {\it Data Conflicts}. 
A specific vertex of a graph may be often associated with a large number of edges, particularly true for skewed graphs. There is the common case that some \hl{vertices} may be updated in conflict by many vertices simultaneously. For preserving the correctness of vertex updating, the specialized mechanisms are presented to enforce the atomicity. For example, for a read-modify-write update of a destination vertex, \cite{1-GRAPHICIONADO,29-Zhou-Highthroughput} propose to use the Content Addressable Memory (CAM) like hardware structure to support finer-granularity memory access, but extra pipeline stalls occur. Similar conflicts can also exist between multiple pipelines. An effective runtime scheduling is expected to avoid these conflicts of vertex updating for high throughput.

\item {\it Workload Balance}. 
Natural graphs in the real world often manifest the power-law distribution\upcite{power-law}. This can result in severe load imbalance \hl{issue} in the sense that a few vertices have extremely high degrees. Workload imbalance may lead to the fact that the loads of some computational logic is overly assigned \hl{while} other light processing units are stalled. More serious is that the loads of the graph computation are often difficult to predict due to the complex data dependencies. An effective runtime scheduling \hl{scheme} for graph processing accelerators should be also expected to dynamically balance hardware resources with even loads for every processing unit as much as possible\upcite{2-EAA}. 

\end{itemize}

\subsection{Communication Model}
The communication model is a well-known pattern that exists commonly to propagate the information between different processing units. We next survey several patterns that have been used in off-the-shelf graph accelerators.

{\it Message-based Pattern.}\quad
Message-based communication model is widely used in distributed environments. In message-based communication model, communication is realized by sending messages among different processing units. These massages can carry the updated data or computation commands that will be execute locally. This model is widely used in HMC-assisted graph processing accelerators\upcite{7-TESSERACT, 12-GRAPHP}. As mentioned previously, the vaults in HMCs communicate with each other via messages. 

Tesseract\upcite{7-TESSERACT} designs the remote function call mechanism via message passing to indicate the running of destination processing cores. The message passing can be used to avoid the cache coherence issues of the processing cores. It can also reduce the atomic operations for shared data. However, a large number of messages come with a high cost of communication time and bandwidth. Partitioning \hl{methods} and coalescing \hl{methods} are usually needed to reduce the number of messages\upcite{12-GRAPHP}. Besides, extra memory copying operations and buffers are also needed.

{\it Shared Memory-based Pattern.}\quad
The shared memory model is suited for the communication between processing units on a single accelerator. The same location of a memory can be accessed and updated by multiple processing units simultaneously. When multiple accelerators are adopted, it is also possible to have a distributed shared memory.

FPGP\upcite{26-FPGP} adopts this model based on FPGAs. It maintains a global shared vertex memory for multiple FPGA boards and each board keeps a vertex cache for multiple processing units. Synchronization between iterations is needed to maintain memory consistency. 
Constrained by limited bandwidth, the global shared vertex memory can limit the scalability of FPGAs. ForeGraph\upcite{30-ForeGraph} uses a distributed shared memory. Shared memory model can usually avoid the redundant copies of graph data and extra storage space in message passing model. It is also easy to implement and design. However, there may exist many data races on the same memory location if some \hl{vertices} are updated by many neighboring vertices.

\subsection{Execution Model}
The execution model typically has two major concerns: \hl{1) scheduling timing, and 2) scheduling order}. The scheduling timing indicates when to execute the vertex programs, which can be often synchronous or asynchronous. The scheduling order indicates the information flow for a vertex program to decide how to update the vertex. They are often used to co-determine when and how a vertex can execute an update if it is active.

{\it Synchronous Mode.}\quad
In the synchronous execution mode, all the vertices in a graph are processed in certain order during each iteration. Between two consecutive iterations, there is a global barrier \hl{to ensure} that all the newly updated vertices in current iteration are visible at the same time in the next iteration for all processors\upcite{Powerswitch}. In graph accelerators, the graph is usually partitioned into subgraphs that are processed by different processing units. When a processing unit finishes its work, it has to wait for other processing units finished. Then the values of different subgraphs are synchronized\upcite{20-CYGRAPH}. During each iteration, only the local values of graph data can be accessed and updated\upcite{26-FPGP}. 

The synchronous execution is easy to realize on graph accelerators and suits for memory-bound graph algorithms. It can utilize the memory bandwidth better because the data is updated in a bulk synchronous way. Many memory accesses can be combined and sequential. However, as discussed before, synchronous mode may require more storage space for local data in each iteration when workloads are unbalanced.

{\it Asynchronous Mode.}\quad
In asynchronous execution mode, each processing unit can start the next iteration immediately when it finishes current workloads. There is no global barrier to synchronize these processing units. 
Asynchronous mode can be used to balance the loads because the processing units are kept busy nearly all the time. This mode suits for the algorithms that converge faster than synchronous execution. Some graph algorithms can only converge under asynchronous execution, e.g., \hl{the} graph coloring algorithm. It also supports dynamic scheduling, e.g., the priority-based scheduling mechanism\upcite{2-EAA} to achieve high performance. However, the disappointing point is that the asynchronous mode requires tremendous efforts to implement on graph accelerators for the sophisticated hardware design\upcite{Ozdal-Architectural}.

{\it Information Flow Directions.}\quad
For executing a vertex program, it is important to decide how to update the value of vertices. \hl{The information flow} between vertices typically \hl{has two kinds of direction}: the push-based mode and the pull-based mode. For an active vertex, the information is propagated from the active vertex to its neighbors in the push mode, while in the pull mode the information is flowed from its neighbors to the active vertex. For BFS algorithm, in the push mode, the values of out-degree neighbors are updated according to active vertices. In the pull mode, the active vertex gets information from its in-degree neighbors to update itself.

Usually, \hl{the} push mode can explicitly select the update vertices but it may cause redundant random accesses when seeking the next frontier. Locks might be needed to ensure the consistency since a vertex may be updated by multiple active vertices. \hl{The pull} mode presents better locality for updated vertices and has natural consistency because the vertices just update themselves. However, it may result in additional overhead for checking whether the updating of neighboring vertices are necessarily executed. 

Push and pull modes can be also combined together and switched at runtime to alleviate the synchronization and communication overhead\upcite{Beamer-HybridBFS}. Ligra\upcite{Ligra} first adopts this method into shared memory graph processing system and Gemini\upcite{Gemini} is the first to apply this hybrid mode to a distributed memory setting which achieves extremely high performance. This hybrid method has also been used in some graph accelerators for performance improvement\upcite{36-DegreeBFS, 18-BetkaouiAPSP}. The switching time is based on the number of active vertices in the frontier and associated unexplored edges. We can switch to the pull mode if the frontier has a high ratio of the unexplored edges for better performance\upcite{36-DegreeBFS}.

\subsection{Scheduling Schemes}
There are many runtime scheduling schemes that can be adopted in graph processing accelerators.

{\it Block-based Scheduling.}
In block-based scheduling, the whole graphs are evenly partitioned into blocks and are distributed to multiple processors. There is no strict order for these partitions to be processed. This scheduling method is widely used for graph processing integrated with multiple accelerators. 

For example, Tesseract\upcite{7-TESSERACT} distributes the graphs to multiple vaults on HMCs to process in parallel. ForeGraph\upcite{30-ForeGraph} partitions the graph into \hl{a} grid and distributes the grid blocks to different FPGA boards. These executions of subgraphs are usually synchronized after each iteration. The batch-based scheduling can easily help achieve massive parallelism among multiple accelerators in a synchronous fashion. However, the workloads of each batch should be balanced to achieve better resources utilization.

{\it Frontier-based Scheduling.}
This kind of scheduling is suitable for those graph algorithms in which only a subset of data needs to be processed in each iteration. A frontier is needed to contain the active data that is to be scheduled. For example, in the vertex-centric model, the frontier contains the active vertices that need to be executed for each iteration. The scheduler gets a vertex from the frontier and checks the state array to decide the data path of the vertex\upcite{4-TUNAO, 17-BetkaouiBFS, Ozdal-Architectural}. \hl{The} frontier-based scheduling can help process most of graph algorithms. However, the frontier might be modified frequently by multiple vertices which contend for updating the same vertex with serious race conditions. The specialized hardware circuit design may be a viable solution for efficiently supporting multiple simultaneous updates.

{\it Priority-based Scheduling.}
In the priority-based scheduling, the scheduled items are assigned a priority flag which represents the execution order. This kind of scheduling is usually combined with the frontier-based approach where the active vertices are ranked. It can also be used to schedule the order of messages to be processed\upcite{7-TESSERACT}. Prioritiy-based scheduling can help some graph algorithms converge faster in a asynchronous execution model, e.g., the PageRank algorithm\upcite{2-EAA}. 

For example, a specialized synchronization unit is designed in \cite{2-EAA} to rank and schedule the active vertex. These active vertices are maintained in an active list, and they are then executed according to the ranking value. However, the newly created dependencies based on the priorities may bring extra synchronization overhead. Fortunately, the latency can usually be compensated by the gains because of the fast convergence. 

{\bf Remarks.}\quad
A single graph processing accelerator may have limited hardware resources and memory capacity. For mobilizing the potentials of these resources, in addition to the effective resource layout, an efficient runtime scheduling scheme is the key, which decides when and where a specified data is supposed to be processed. Considering the complexity of the hardware circuit layouts, unlike the pure software implementations, the runtime scheduling on a graph accelerator has to be co-designed with the necessary hardware supports in many cases for better efficiency.

For instance, software-assisted runtime scheduling for ensuring the sequential consistency can use locking mechanisms that are easy to implement. \hl{However, these mechanisms} can be also error-prone and even suffer from significant performance degradation in hardware implementation. The specialized hardware \hl{supports} with CAM structure\upcite{FPGACAM} or more advanced designs\upcite{40-YaoPACT} make the scheduling for sequential consistency easy. Runtime scheduler can therefore focus more on the parallelism exploitation\upcite{Ozdal-Architectural}. In addition, this also greatly mitigates the atomicity overhead. Combined with irregular accesses and large sizes of graphs, more extra efforts still have to be done for runtime scheduling. 

\section{Graph Accelerator Evaluation}

The key issues of the design and implementation of graph accelerators have been summarized in previous sections. These designs differ in preprocessing methods, programming models, and hardware architectures. Here we summarize the key metrics in existing work and make a detailed discussion from following aspects.

\begin{itemize}[leftmargin=*]

\item \textit{Evaluation Metrics}.  Evaluation metrics presented in this paper include the typical design techniques, hardware platform parameters, performance metrics, and energy efficiency metrics. These metrics provide an overall view of different graph accelerators. 
 
\item \textit{Summary of Results}. Based on the evaluation metrics, we analyze these results and make a discussion from five aspects: graph benchmarks, platform parameters, preprocessing, graph processing frameworks and programming models. Various kinds of graph benchmarks and platforms make it difficult for a fair comparison of different accelerators. Different kinds of design methods can also influence the performance. We argue that it demands standard graph accelerator benchmarks for efficient evaluations.

\item \textit{A Case Study}. In the review, we find that there is no absolute winner among existing graph processing accelerators in terms of performance and energy efficiency. In this section, we choose another angle to study the design and implementation of a state-of-the-art accelerator\upcite{40-YaoPACT} in more depth so that readers can have a more in-depth understanding on the three core components. 

\end{itemize}

\setcounter{table}{3}
\tabcolsep 9pt
\renewcommand\arraystretch{1.3}
\begin{table*}[!htb]
\centering
\caption{\label{t1} Overview of Graph Processing Accelerators}\vspace{-2mm}
{\footnotesize
\begin{tabular*}{\linewidth}%
{lp{2.0cm}<{\centering}p{1.3cm}<{\centering}p{1cm}<{\centering}p{1cm}<{\centering}p{1.5cm}<{\centering}p{1cm}<{\centering}p{0.5cm}<{\centering}p{2cm}<{\centering}l}
\hline\hline\hline
Year & System         & Architecture & Data Layout & Prepro\-cessing & Programming Model & Generality  &Sche\-duling      & Compared System & ID \\
\hline
2016 & Graphicionado\upcite{1-GRAPHICIONADO}  & ASIC         & COO         & Y             & V/Sync            & Various &  F        & GraphMat\upcite{GraphMat}       & 1     \\
2016 & EEA\upcite{2-EAA}            & ASIC         & CSR         & Y             & V/Async           & Various &P          & Host            & 2     \\
2017 & TuNao\upcite{4-TUNAO}          & ASIC         & COO         & Y             & V/Async           & Various   & F        & Cusha\upcite{Cusha}          & 3     \\
2017 & GAA\upcite{5-GAA}           & ASIC         & CSR         & Y             & V/Async           & Various    & P       & Host            & 4     \\
2018 & Ozdal \textit{et~al.}\upcite{6-TEMPLATEGAA}  & ASIC         & CSR         & Y             & V/Async           & Various   & P        & GAP\hl{\upcite{GAP}}            & 5     \\
2015 & Tesseract\upcite{7-TESSERACT}     & PIM          & -           & Y             & V/Sync            & Various   & B        & Host            & 6     \\
2017 & GraphPIM\upcite{8-GRAPHPIM}      & PIM          & CSR         & N             & V/Sync            & Various  & F         & GraphBIG\upcite{GraphBIG}        & 7     \\
2017 & RPBFS\upcite{9-RPBFS17}         & PIM          & CSR         & Y             & -/Sync            & BFS      & B         & Enterprise\upcite{Enterprise}      & 8     \\
2018 & GraphR\upcite{10-GRAPHR}        & PIM          & COO         & Y             & E/Sync            & Various   & B        & GridGraph\upcite{Gridgraph}       & 9     \\
2018 & RPBFS\upcite{11-RPBFS18}         & PIM          & CSR         & Y             & -/Sync            & BFS     & B          & Enterprise\upcite{Enterprise}      & 10    \\
2018 & GraphP\upcite{12-GRAPHP}        & PIM          & -           & Y             & V/Sync            & Various   & B        & Tesseract\upcite{7-TESSERACT}      & 11    \\
2018 & GraphH\upcite{13-GRAPHH}        & PIM          & COO         & Y             & E/Sync            & Various  & B         & Tesseract\upcite{7-TESSERACT}       & 12    \\
2010 & Wang \textit{et~al.}\upcite{15-WANG-MESSEGEBFS}     & FPGA+SoC     & CSR         & Y             & V/Sync            & BFS  & F             & Cell BE\upcite{Cell-BE}         & 13    \\
2011 & Betkaoui \textit{et~al.}\upcite{16-GRAPHLET} & FPGA         & CSR         & N             & V/Sync            & GC  & B      & GraphCrunch\upcite{GraphCrunch}    & 14    \\
2012 & Betkaoui \textit{et~al.}\upcite{17-BetkaouiBFS} & FPGA         & CSR         & N             & V/Sync            & BFS  & B             & PACT11\upcite{PACT11}          & 15    \\
2012 & Betkaoui \textit{et~al.}\upcite{18-BetkaouiAPSP} & FPGA         & CSR         & N             & V/Sync            & APSP   & B           & HPCC11\upcite{HPCC11}         & 16    \\
2014 & GraphGen\upcite{19-GRAPHGEN}      & FPGA         & COO         & Y             & V/Sync            & Various & F          & Host            & 17    \\
2014 & CyGraph\upcite{20-CYGRAPH}       & FPGA         & CUST        & Y             & V/Sync            & BFS     & F          & ASAP12\upcite{17-BetkaouiBFS}          & 18    \\
2015 & Attia \textit{et~al.}\upcite{21-AttiaAPSP}   & FPGA         & CUST        & Y             & V/Sync            & APSP & F             & BGL\upcite{BGL}        & 19    \\
2015 & Umuroglu \textit{et~al.}\upcite{22-UmurogluHybridBFS} & FPGA+SoC     & CSC         & Y             & -/Sync            & BFS  & F             & Host            & 20    \\
2015 & Zhou \textit{et~al.}\upcite{23-ZhouSSSP}    & FPGA         & COO         & Y             & E/Sync            & SSSP   & B           & CyGraph\upcite{20-CYGRAPH}         & 21    \\
2015 & Zhou \textit{et~al.}\upcite{24-ZhouPAGERANK}    & FPGA         & COO         & Y             & E/Sync            & PageRank  & B        & Host            & 22    \\
2015 & GraphSoC\upcite{25-GraphSoC}      & FPGA+SoC     & -           & Y             & V/Sync            & Various & B          & Host            & 23    \\
2016 & FPGP\upcite{26-FPGP}          & FPGA         & COO         & Y             & V/Sync            & BFS    & B           & GraphChi\upcite{GraphChi}       & 24    \\
2016 & GraVF\upcite{27-GraVF}         & FPGA         & -           & Y             & V/Sync            & various  & B         & -               & 25    \\
2016 & GraphOps\upcite{28-GraphOps}      & FPGA         & CUST        & Y             & V/Sync            & Various & F          & X-Stream\upcite{X-Stream}       & 26    \\
2016 & Zhou \textit{et~al.}\upcite{29-Zhou-Highthroughput}    & FPGA         & COO         & Y             & E/Sync            & various & B          & X-Stream\upcite{X-Stream}       & 27    \\
2016 & SpMV\upcite{3-NOVELSPMV}          & FPGA         & -           & N             & -/Sync            & SpMV   & B           & Host            & 28    \\
2017 & ForeGraph\upcite{30-ForeGraph}      & FPGA         & COO         & Y             & E/Sync            & various  & B         & FPGP\upcite{26-FPGP}           & 29    \\
2017 & Ma \textit{et~al.}\hl{\upcite{31-FPGAHTM}}       & FPGA         & -           & N             & -/Async           & various   & B        & Host            & 30    \\
2017 & Zhang \textit{et~al.}\upcite{32-FPGAHMCBFS}   & FPGA         & CSR         & Y             & V/Sync            & BFS    & F           & FPGP\upcite{26-FPGP}            & 31    \\
2017 & Zhou \textit{et~al.}\upcite{34-ZhouCPUFPGA}    & FPGA+CPU     & CUST        & Y             & Hybrid/Sync       & Various  & F         & GraphMat\upcite{GraphMat}        & 32    \\
2018 & Zhang \textit{et~al.}\upcite{36-DegreeBFS}   & FPGA         & CSR         & Y             & V/Sync            & BFS    & F           & FPGA17\upcite{32-FPGAHMCBFS}         & 33    \\
2018 & Khoram \textit{et~al.}\upcite{37-khoram-FPGAHMC}   & FPGA+HMC     & CSR         & Y             & V/Sync            & BFS  &F             & FPGA17\upcite{32-FPGAHMCBFS}         & 34    \\
2018 & FASTCF\upcite{38-FASTCF}        & FPGA         & COO         & Y             & E/Sync            & CF    & B            & SIGMOD14\upcite{Satish-Navigating}        & 35    \\
2018 & Yao \textit{et~al.}\upcite{40-YaoPACT}     & FPGA         & CSR/CSC     & Y             & V/Sync            & Various & F           & ForeGraph\upcite{30-ForeGraph}      & 36    \\
2018 & GraFBoost\upcite{41-GRAFBOOST}      & FPGA+Flash         & CSR         & Y             & E/Sync            & Various  & B         & FlashGraph\upcite{FlashGraph}     & 37  
\\\hline\hline\hline
\end{tabular*}
}
\end{table*}
\baselineskip=13.2pt plus.2pt minus.2pt
\parskip=0pt plus.2pt minus0.2pt

\setcounter{table}{4}
\tabcolsep 9pt
\renewcommand\arraystretch{1.3}
\begin{table*}[!htb]
\centering
\caption{\label{t2} Parameters of Graph Accelerator Platforms}\vspace{-2mm}
{\footnotesize
\begin{tabular*}{\linewidth}{ccp{1.5cm}<{\centering}p{2.5cm}<{\centering}cccc}\hline\hline\hline
ID & Compute Device               & Frequency  & On-chip Memory & Off-chip Memory & Bandwidth & Method\\
\hline
1  & Streams*8                    & 1GHz       & eDRAM 64MB     & DDR4*4          & 68GB/s   & S \\
2  & AU*4                         & 1GHz       & Cache 34.8KB   & DDR4            & 12.8GB/s & S \\
3  & ECGRA                        & 300MHz     & Cache 2.4MB    & -               & 288GB/s  & M \\
4  & AU*4                         & 1GHz       & -              & DDR4            & 12.8GB/s & S  \\
5  & AU*4                         & 1GHz       & -              & DDR4            & 12.8GB/s & S \\
6  & HMC (512cores)               & 2GHz       & Cache 16MB     & HMC1.0*16       & 8TB/s    & S \\
7  & CPU (16cores)                & 2GHz       & Cache 16MB     & HMC2.0          & 480GB/s  & S \\
8  & ReRAM (1024*1024)            & 1.2GHz     & eDRAM 4MB      & ReRAM           & 50GB/s   & S \\
9  & ReRAM (32*64)                & -          & ReRAM          & Disk            & -        & S \\
10 & ReRAM (1024*1024)            & 1.2GHz     & eDRAM 4MB      & ReRAM           & 50GB/s   & S \\
11 & HMC (512cores)               & 1GHz       & Cache 49MB     & HMC2.1*16       & 5TB/s    & S \\
12 & HMC (512cores)               & 1GHz       & SRAM 576MB     & HMC2.1*16       & 5TB/s    & S \\
13 & Virtex-5 FPGA                & 100MHz     & BRAM 1.29MB    & DDR3            & 0.1GB/s  & S \\
14 & Virtex-5 FPGA*4              & 75MHz      & BRAM 5.18MB    & -               & 80GB/s   & M \\
15 & Virtex-5 FPGA*4              & 75MHz      & BRAM 5.18MB    & -               & 80GB/s   & M \\
16 & Virtex-5 FPGA*4              & 75MHz      & BRAM 5.18MB    & -               & 80GB/s   & M \\
17 & Virtex-6 FPGA                & 100MHz     & BRAM 1.87MB    & DDR2            & 6.4GB/s  & M \\
18 & Virtex-5 FPGA*4              & 75MHz      & BRAM 5.18MB    & -               & 80GB/s   & M \\
19 & Virtex-5 FPGA*4              & 75MHz      & BRAM 5.18MB    & -               & 80GB/s   & M \\
20 & FPGA/ARM                     & 150/666MHz & BRAM 0.56MB    & DDR3            & 3.2GB/s  & M \\
21 & Virtex-7 FPGA                & 200MHz     & BRAM 4.5MB     & DDR3            & 20GB/s   & M \\
22 & Virtex-7 FPGA                & 200MHz     & BRAM 8.375MB   & DDR3            & 20GB/s   & S \\
23 & ZC706 FPGA/SoC               & 250MHz     & BRAM 70KB      & DDR3            & -        & M \\
24 & Virtex-7 FPGA                & 100MHz     & BRAM 4.76MB    & DDR3            & 12.8GB/s & M \\
25 & Virtex-7 FPGA                & 150MHz     & BRAM 6.6MB     & DDR3            & -        & M \\
26 & Virtex-6 FPGA                & 150MHz     & BRAM 4.76MB    & DDR3            & 38.4GB/s & M \\
27 & Virtex UltraScale FPGA       & 250MHz     & BRAM 12.8MB    & DDR4            & 19.2GB/s & S \\
28 & FPGA*4                       & -          & -              & DDR3*8          & 102.4GB/S & M \\
29 & Virtex UltraScale FPGA       & 200MHz     & BRAM 16.61MB   & DDR4            & 19.2GB/s & S \\
30 & Virtex UltraScale 440 FPGA*2 & 200MHz     & BRAM 22MB      & DDR3            & 51.2GB/s & S \\
31 & AC-510 FPGA                  & 125MHz     & BRAM 4.75MB    & HMC2.0          & 60GB/s   & M/S \\
32 & Arria10 FPGA/ Xeon-cores*14  & -          & BRAM 6.6MB     & DDR3            & 12.8GB/s & M \\
33 & AC-510 FPGA                  & 125MHz     & BRAM 4.75MB    & HMC2.0          & 60GB/s   & M/S \\
34 & AC-510 FPGA                  & 125MHz     & BRAM 4.75MB    & HMC2.0          & 60GB/s   & M \\
35 & Virtex UltraScale+ FPGA      & 150MHz     & RAM 43.3MB     & DDR4*2          & 38.4GB/s & M \\
36 & Virtex Ultrascale+ FPGA      & 250MHz     & BRAM 9.49MB    & DDR4            & 19.2GB/s & S \\
37 & VC707 FPGA/Flash             & 125MHz     & BRAM 4MB       & DDR3            & 10GB/s  & M
\\\hline\hline\hline
\end{tabular*}
}
\end{table*}
\baselineskip=13.2pt plus.2pt minus.2pt
\parskip=0pt plus.2pt minus0.2pt

\setcounter{table}{5}
\tabcolsep 9pt
\renewcommand\arraystretch{1.3}
\begin{table*}[!htb]
\centering
\caption{\label{t3} Comparison of Performance and Energy Efficiency}\vspace{-2mm}
{\footnotesize
\begin{tabular*}{\linewidth}%
{p{0.3cm}<{\centering}p{1.6cm}<{\centering}p{1.5cm}<{\centering}p{1.4cm}<{\centering}p{1.1cm}<{\centering}p{1.5cm}<{\centering}ccc}
\hline\hline\hline
ID & BFS & SSSP  & PageRank  & SpMV     & Energy & $|V|_{max}$ & $|E|_{max}$          & Datasets    \\
  & (GTEPS) & (GTEPS) & (GTEPS) & (GTEPS)  &  Efficiency &  (Million) & (Million)  & Type \\
\hline
1  & 0.125$\sim$2.6   & 0.25$\sim$2.3    & 4.5$\sim$4.75    & -        & 7W                            &  61.57  &  1468.36  & Social/RMAT      \\
2  & -           & SP          & SP          & -        & 3.375W                        &  67     &  1000        & Social/Kronecker \\
3  & SP          & SP          & SP          & SP       & 9.6W                          &  7.4    &  192        & Social           \\
4  & -           & SP          & SP          & -        & SP                            &  67     &  1000        & Social/Kronecker \\
5  & -           & SP          & SP          & -        & SP                            &  16.8   &  268       & Social/Kronecker \\
6  & -           & SP          & SP          & -        & 94 mW/mm2                     &  7.4    &  194        & Social           \\
7  & SP          & SP          & SP          & -        & -                             &  1      &  28.8         & LDBC             \\
8  & 0.2$\sim$1.2     & -           & -           & -        & -                             &  2.39   &  7.6       & Social           \\
9  & SP          & SP          & SP          & SP       & 1.08pJ(r) 3.91nJ(w)           &  4.8    &  106        & Social           \\
10 & 0.2$\sim$1.2     & -           & -           & -        & 1.59pJ(r) 5.53nJ(w)           &  1.96   &  5.53      & Social           \\
11 & SP          & SP          & SP          & -        & SP                            &  4.8    &  6.9        & Social           \\
12 & SP          & -           & 320$\sim$350     & -        & 133 mW/mm$^2$                    &  41.7   &  6640      & Social           \\
13 & 0.16$\sim$0.79   & -           & -           & -        & -                             &  0.064  &  1.024    & Synthetic        \\
14 & -           & -           & -           & -        & -                             &  0.3    &  3          & LEDA             \\
15 & 0.25$\sim$2.6    & -           & -           & -        & -                             &  16     &  1024        & RMAT             \\
16 & -           & -           & -           & -        & -                             &  0.038  &  -        & fMRI             \\
17 & -           & -           & -           & -        & -                             &  0.11   &  0.34      & Image               \\
18 & 1.68$\sim$2.2    & -           & -           & -        & -                             &  8      &  512          & RMAT             \\
19 & -           & -           & -           & -        & -                             &  0.065  &  4.19     & RMAT             \\
20 & 0.09$\sim$0.255  & -           & -           & -        & -                             &  2      &  67           & RMAT             \\
21 & -           & 1.6         & -           & -        & -                             &  1      &  -            & RMAT             \\
22 & -           & -           & 0.27$\sim$0.38   & -        & -                             &  2.39   &  7.6       & Social           \\
23 & -           & -           & -           & 0.015    & -                             &  0.017  &  0.126    & SpMV             \\
24 & 0.01$\sim$0.012  & -           & -           & -        & -                             &  1400   &  6600         & Social           \\
25 & 3.5         & -           & 3           & -        & -                             &  0.0025 &  0.01                  & Synthetic                \\
26 & -           & -           & 0.035$\sim$0.115 & 0.2$\sim$0.75 & -                             &  2.39   &  30.6      & Social           \\
27 & -           & 0.657$\sim$0.872 & -           & -        & 19.06$\sim$24.22W                 &  4.7    &  65.8       & Social           \\
28 & -           & -           & -           & 0.316    & 2 MTEPS/W                     & -       & -            & -        \\
29 & 0.897$\sim$1.458 & -           & 0.997$\sim$1.856 & -        & -                             &  1410   &  6640        & Social           \\
30 & SP          & SP          & -           & -        & 5$\sim$8W                          &  24     &  64          & Synthetic        \\
31 & 0.13$\sim$0.166  & -           & -           & -        & -                             &  33.5   &  536.8     & RMAT             \\
32 & 0.33$\sim$0.67   & 0.063$\sim$0.075 & -           & -        & -                             &  10     &  160         & RMAT             \\
33 & 0.4$\sim$152.6   & -           & -           & -        & 43.6W                         &  23.9   &  577.1     & Social/RMAT      \\
34 & 0.1$\sim$0.65    & -           & -           & -        & -                             &  16     &  252.8                  & Social           \\
35 & -           & -           & -           & -        & 13.8W                         &  1.3    &  460        & Bipartite        \\
36 & 1.5$\sim$3.5     & -           & 1.25$\sim$2.5    & -        & -                             &  3.07   &  117       & Social           \\
37 & 0.057$\sim$0.075 & -           & SP          & -        & 50W                           &  3000   &  128000    & Social/Kronecker
\\\hline\hline\hline
\end{tabular*}
}
\end{table*}

\baselineskip=18pt plus.2pt minus.2pt
\parskip=0pt plus.2pt minus0.2pt

\setcounter{figure}{3}
\begin{figure*}[!htb]
\centering
\subfigure[Relationship of energy efficiency and graph size]{\includegraphics[width=8cm]{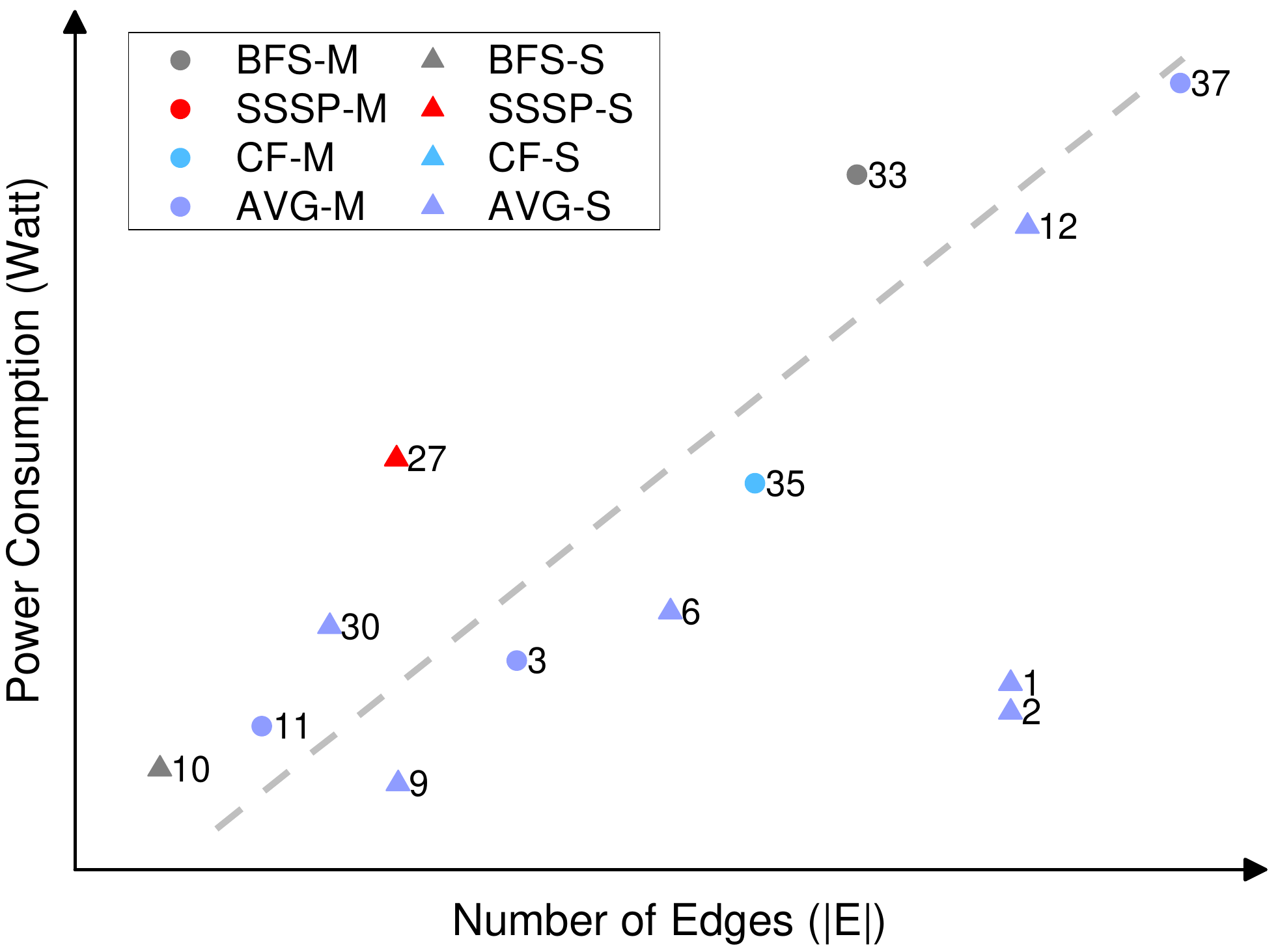}}
\hspace{1cm}
\subfigure[Relationship of performance and graph size]{\includegraphics[width=8cm]{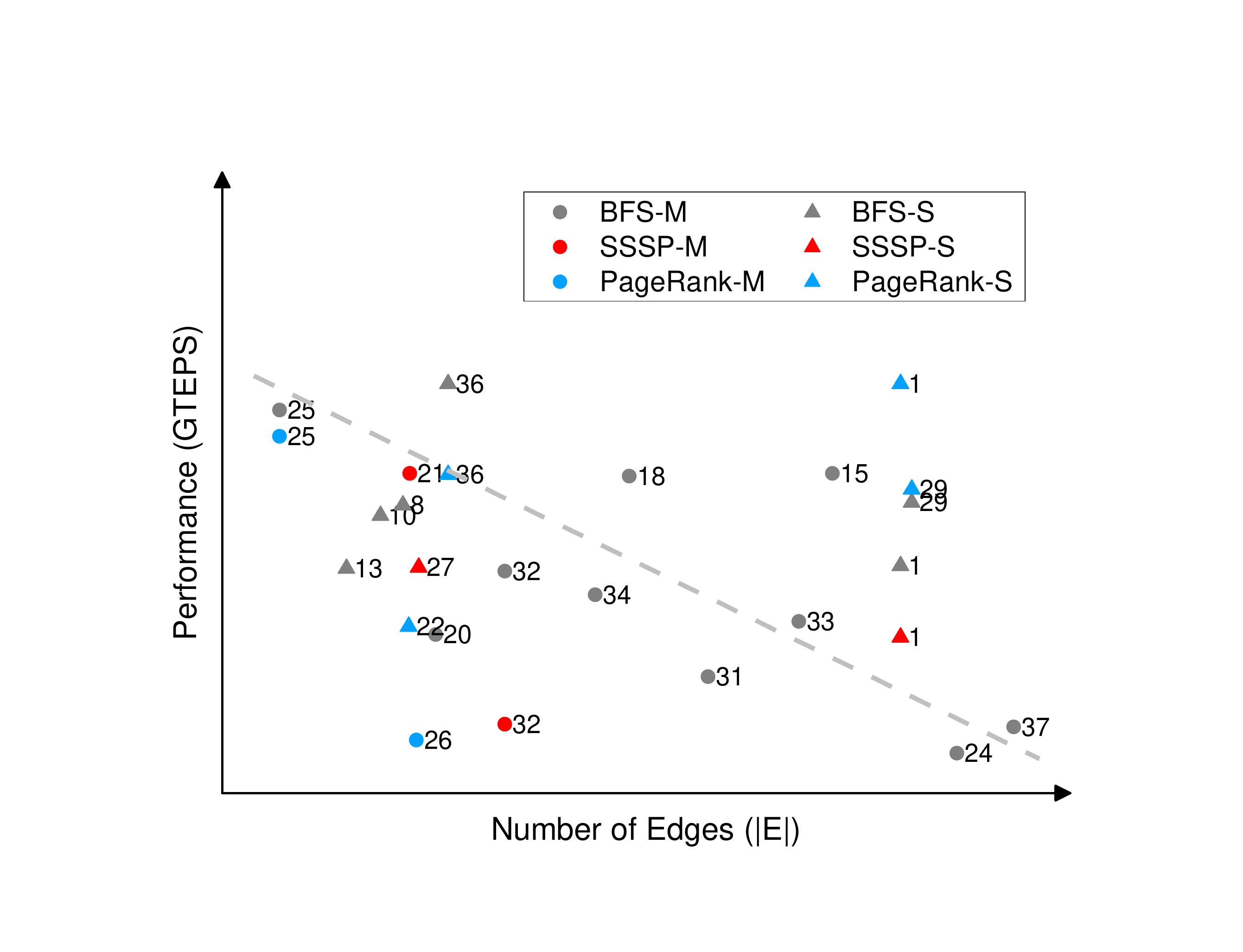}}
\caption{The relative development trend of (energy efficiency and/or performance) results for existing state-of-the-art graph processing accelerators, and explicit results can refer to Table 6 for details.``-M'' represents the measurement-based results and``-S'' represents the simulation-based results.}
\end{figure*}
\baselineskip=18pt plus.2pt minus.2pt
\parskip=0pt plus.2pt minus0.2pt

\subsection{Evaluation Metrics}
In order to assess the graph accelerators, existing work typically uses TEPS as the performance metric, TEPS/W or power consumption \hl{(watt, or 
joule per read/write)} as energy efficiency metric, respectively. These metrics basically give an overall 
evaluation of the graph acceleration system.

Key parameters of existing graph accelerators for evaluation are divided into three aspects. Table 4 gives an overview of a graph processing accelerator including the pre-processing, programming models, and compared systems. Note that each study is assigned with a unique ID which is also used for the same accelerator. Table 5 summarizes the hardware parameters of graph accelerators. Table 6 summarizes the comparison of performance and energy efficiency reported in the related work.

For fidelity, the label \hl{``M''} and \hl{``S''} are used to distinguish the measurement-based results and simulation-based results respectively in Table 5.
We try to provide the actual performance/energy metrics, but 
some related work has only the relative performance/energy over the compared systems. 
We thus cannot infer the actual accelerator performance according to their original results.
In this case, the performance/energy is labeled as \hl{``SP''} (Speedup) in Table 6.
Some accelerators support only a single graph algorithm or a few graph algorithms.
The corresponding performance 
will be labeled as ``-''. In addition, we use abbreviations for some long terminologies
because of the limited space. In programming model category, we use \hl{``V'', ``E''} to 
represent the vertex-centric model and edge-centric model, respectively. 
When the model is not clearly named, we use ``-'' instead. Similarly, 
we use \hl{``Sync'', ``Async''} to represent the synchronous execution and asynchronous 
execution, respectively. Block-, frontier- and priority-based scheduling methods are represented by \hl{``B'', ``F'', and ``P''}, respectively.

\subsection{Summary of Results}
We analyze the summary in the following aspects, including graph benchmark, platform parameter, preprocessing, graph processing framework, programming models and runtime scheduling.\\
\indent{\it 1) Graph Benchmark.} When comparing the accelerators, the benchmark is of vital importance to understand the effectiveness of the design and implementation of a graph processing accelerator. A graph benchmark 
consists of at least four aspects including graph layouts, types of input graphs, 
size of the graphs, and graph algorithms. As shown in Table 4, graph layouts are different across the existing studies on graph processing accelerators.  Thus, in fact it requires further research for developing a fair and practical benchmark for evaluating different graph processing accelerators. Particularly, we have the following observations for further research.

First, existing studies use different storage \hl{layouts}. Some of them adopt the edge list, some of them use CSR/CSC, and some of them utilize the customized 
layout (CUST). They affect the memory access patterns dramatically and the performance accordingly. 

Second, according to Table 6, the types of the graphs used in the accelerators are not totally the same. Types of graphs used in prior work including real-world graph, i.e., social network graph, road network graph, \hl{and functional magnetic resonance imaging (fMRI) graphs}. There are also synthetic graphs, i.e., \hl{the recursive matrix (RMAT) graph, the Kronecker graph, the graphs generated by the Linked Data Benchmark Council (LDBC), and the graphs generated by the Library of Efficient Data Types and Algorithms (LEDA)}. Different combinations lead to diverse results.

Third, graph algorithms used in different graph accelerator designs are also usually different. 
If the algorithms used are different, comparing 
the metrics of performance and energy efficiency needs to be improvable and justified. 

Fourth, graph size is another key graph parameter, but it is not sufficiently considered in previous work.
The graph size used in different graph accelerators varies in a large range as the maximum number of vertices and edges presented in Table 6. 
Some graphs have less than a million vertices while some of them have more than a billion. 
Given even the same type of graph algorithms, the graphs \hl{can} involve different \hl{sizes}, especially the RMAT graphs. \hl{The number of vertices or edges may vary according to the configuration of the graph generator. As a result, different average degrees of graphs can result in distinct parallelism and data locality of vertices.}
Therefore, this may lead to different performance in the end.\\
\indent{\it 2) Platform Parameter.}
We find that, even with the same hardware component design, existing graph processing accelerators have different parameter settings. 
According to Table 5, it is clear that the platforms, i.e., ASIC, PIM and FPGA used in different accelerator 
designs, make a big difference on the resulting performance and energy efficiency.
This is expected since the implementation frequency may already be different in an order of magnitude.

However, the parameters of \hl{the same kind of platform} also vary dramatically. For instance, the largest FPGA on-chip 
memory is around 44MB while the smallest one is only 0.25MB. Similarly, the memory \hl{bandwidths} of the same 
type of platforms also differ significantly. Large memory bandwidth allows more parallel processing. Large on-chip memory improves the memory access efficiency. The platform parameters 
can have considerable influence on performance and energy efficiency.

{\it 3) Preprocessing.}
As discussed in Section 3, preprocessing is usually beneficial to graph processing as it improves 
the data locality or memory access patterns. While we notice that some graph processing 
accelerators do not involve preprocessing at all, it is unfair to make \hl{an} end-to-end comparison 
to the ones with preprocessing. In addition, the accelerators with preprocessing can also have 
diverse preprocessing efforts. When the preprocessing efforts are different, it is also tricky 
to compare the accelerators. In some of the occasions, when the preprocessing cost can be fully amortized, 
we may just ignore the preprocessing overhead. It may not be the case when the application is sensitive to preprocessing 
cost as suggested in \cite{Malicevic-Everything}.

{\it 4) Graph Processing Framework.} 
According to the ``Generality'' column in Table 4, most of the graph processing accelerators 
target a set of typical graph processing algorithms, while the other accelerators may focus on 
optimizing a specific graph processing algorithm. It is essentially a trade-off between 
generality and performance. It is not fair to compare these accelerators when the 
``Generality'' is different.

{\it 5) Programming Model.}
From the tables, it can be found that different programming models are used in the graph processing accelerators.
\hl{The accelerators can be implemented in either the synchronous model or the asynchronous model}. Also, some accelerators follow a 
vertex-centric processing model \hl{while others} choose the edge-centric model. Note that there is also 
one graph accelerator based on the hybrid model. Different models may also influence the performance 
of graph accelerators. Nevertheless, there is no clear difference in terms of the ease of 
programming. Different from the above parameters, accelerators with different programming models 
remain comparable. 

{\it 6) Development Trend.}
For further exploration of the results, Fig.4 make a qualitative analysis of the relative development trend. These two charts only present the relative position of the results for a quick evaluation. More explicit details can refer to Table 6.

\hl{Fig.4(a)} depicts the relative energy efficiency (represented in power consumption) of investigated graph processing accelerators as \hl{the} graph size increases. \hl{Fig.4(b)} illustrates the relative performance of \hl{the} investigated graph processing accelerators for BFS, SSSP and PageRank with different graph sizes. The graph size is measured by the largest number of edges in respective literature because the number of edges is usually much larger than \hl{the number of} vertices in the datasets. Edge numbers are depicted in the format of offset reciporcal. The power consumption and performance are depicted in a logit format for qualitative comparison. The ID \hl{number} of \hl{each} graph processing accelerator is labeled besides \hl{correspond accelerator's} data point \hl{in Fig.4}. Note that all the data are based on the explicit descriptions in relevant literatures, and the measurement-based results are distinguished from simulation-based results for the fidelity.

Power consumption is an important metric to measure the energy efficiency\upcite{2-EAA}. The power consumption in \hl{Fig.4(a)} presents an increasing trend as the graph size increases. This is because that it generally demands more computing and storage resources to handle large graphs. Besides, different kinds of hardware designs can contribute to various energy behaviours. The accelerator with the lowest power consumption adopts the emerging ReRAM which \hl{has} intuitive high energy efficiency\upcite{10-GRAPHR}. In order to process larger graphs, the hosts may be involved and result in higher power consumption\upcite{41-GRAFBOOST}. In \hl{Fig.4(a)}, accelerators with IDs by $1$\upcite{1-GRAPHICIONADO} and $2$\upcite{2-EAA} can handle large graphs with good energy efficiency, which are both ASIC-based accelerators. This is because of the dedicated circuit designs and memory subsystems.

As for performance analysis, in spite that the results vary in different accelerators, the results show that the performance acts in a descend trend with graph size \hl{increasing}. This is obvious for \hl{the} BFS algorithm. Note that for \hl{the} SSSP and PageRank algorithms, there is a lack of explicit evaluation results in existing literatures and only limited data points are depicted in \hl{Fig.4(b)}. Most of the results with high performance are based on \hl{small graph size that the graph can fit into the on-chip/board memories}. However, with graph size \hl{increasing}, performance based on single accelerator decreases because external storages are often required\upcite{26-FPGP, 41-GRAFBOOST}. Some designs based on multiple accelerators can maintain high performance when deal with large graphs\upcite{17-BetkaouiBFS, 30-ForeGraph} because the graphs can still be held in on-chip/board memory.

{\bf Remarks.}\quad It gets clear that comparing different graph accelerators is extremely challenging
due to the distinct evaluation parameters. To resolve this problem, the common practice 
in prior work is to \hl{compare the accelerator with} some known systems as shown in Table 4. However, the 
compared systems used in different accelerators are still not comparable. For example, different accelerators adopt various strategies in preprocessing, parallel graph computation models and runtime scheduling schemes. As a result, 
the accelerator evaluation and peer comparison are still trapped into a deadlock.
We conjecture that the lack of graph accelerator benchmarks and reference designs is the root of this problem. 
To this point, developing an open-sourced benchmark as well as an easy-to-port reference design can be 
a potential solution to make a fair evaluation.

\setcounter{figure}{4}
\begin{figure*}[!htb]
\centering
  \includegraphics[scale=0.6]{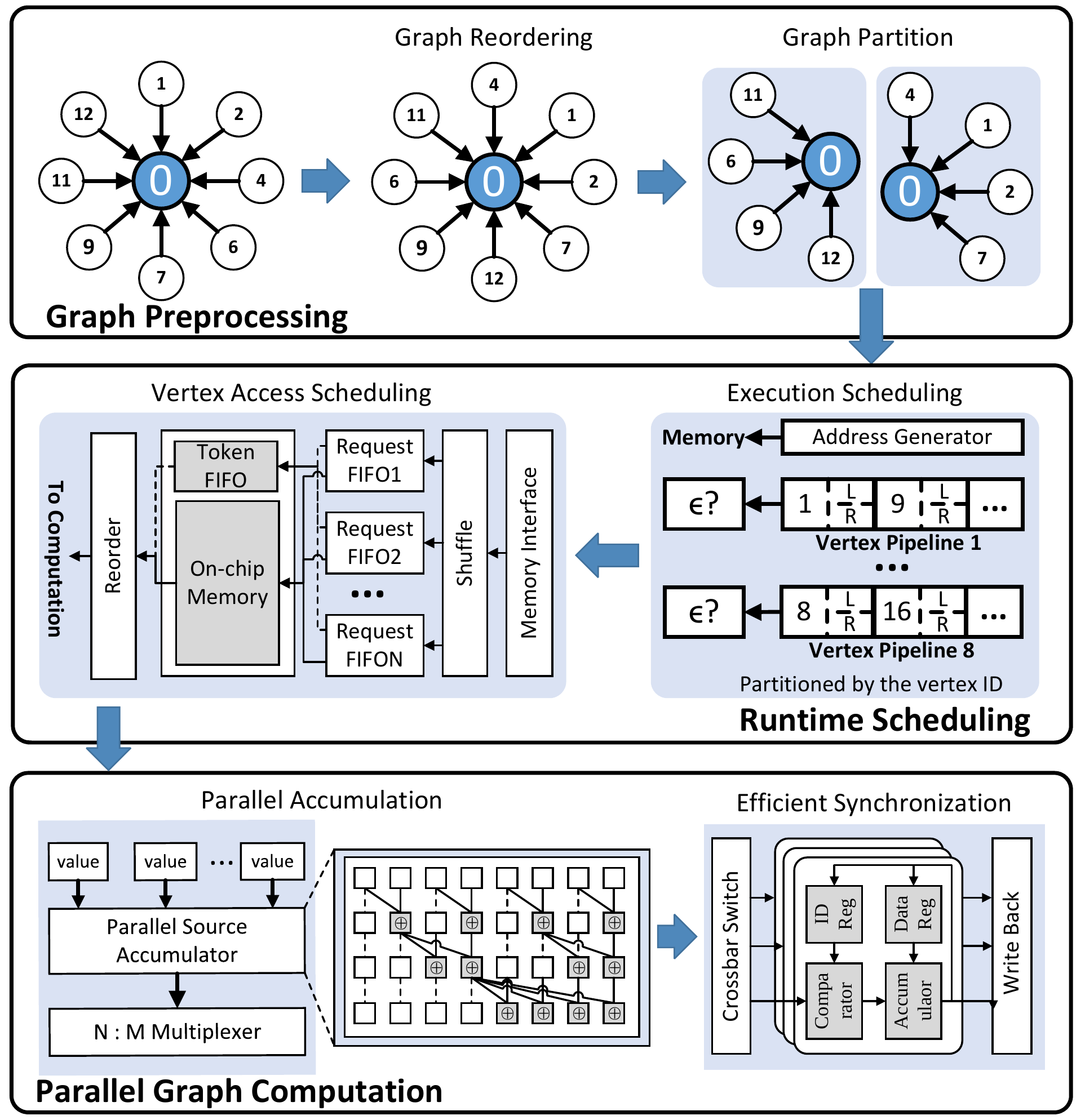}\\
  \caption{The workflow decomposition of AccuGraph in accordance with three major components (described in Fig.1) of preprocessing, parallel graph computation \hl{and} runtime scheduling}
\end{figure*}
\baselineskip=18pt plus.2pt minus.2pt
\parskip=0pt plus.2pt minus0.2pt

\subsection{A Case Study: AccuGraph\upcite{40-YaoPACT}}

As a representative state-of-the-art FPGA-based graph processing accelerator, AccuGraph\upcite{40-YaoPACT} has achieved impressive performance results with the dedicated hardware design for parallelizing the vertex updates that involve conflicts. For better understanding this survey, Fig.5 re-decomposes the original workflow of AccuGraph as a case study according to different stages that we have identified previously.

{\it Preprocessing.}\quad For saving the space of on-chip memories, AccuGraph follows to use the compact graph representation with CSR. In an effort to balance the number of vertex accesses, AccuGraph presents an index-aware ordering to reorder the edges of each vertex by following a simple hash function of MOD($n$) where $n$ is up to the number of on-chip subgraph partitions. As for graph partition, \hl{considering} that AccuGraph uses a pull-based model for high-throughput pipeline design, a vertex-cut graph partitioning method is used \hl{to ensure} the sequential access of the in-degree edges of each vertex. 

{\it Parallel Graph Computation.}\quad
AccuGraph is built upon a Xilinx Virtex Ultrascale+ FPGA board. In order to avoid the half-bandwidth wasting problem of edge-centric programming model that simultaneously accesses both source and destination vertices, AccuGraph uses the vertex-centric programming model to access source vertices only for ensuring the sequential access of edges. 

The core design of AccuGraph lies in a parallel accumulator with dedicated hardware circuits that can support the simultaneous update of conflicting \hl{vertices}. The key insight is that the atomic operations of many graph algorithms manifest incremental and simplex features, which enables to execute massive conflicting vertex updates in an accumulative fashion.
By handling atomic operations simultaneously and merging their results in parallel, the update operations for the same vertex can be therefore parallelized while preserving the correctness of final results. 

It is also observed that a significant amount of locality exists for accessing associated edges of a particular active vertex. In order to further reduce the synchronization overhead of high-degree vertices, AccuGraph follows to use Copy-on-Write philosophy\upcite{COW} to delay the writeback of vertex data. All intermediately-updated vertex data is stored into a specially designed scratchpad memory. If and only if all associated edges are finished, the updated value of \hl{a} given vertex can be written into the main memory.  

{\it Runtime Scheduling.}\quad
To better leverage the limited number of pins of parallel accumulator, AccuGraph uses an improved frontier-based scheduling. \hl{In} the aspect of computational scheduling, it separately handles the pipelines of vertices and edges for reducing the out-of-order memory accesses. \hl{The edge pipelines access} each edge sequentially while each edge pipeline dynamically adjusts the number of vertices to be processed via a degree-aware scheduling mechanism. As for memory access scheduling, the sparsity of graph often leads to the imbalance of accessing vertices. AccuGraph significantly enhances the throughput of on-chip computation by presenting an out-of-order approach for accessing the value of vertex.


\section{Challenges and Opportunities}
With the recent efforts, graph processing accelerators have experienced a series of significant technical advances for achieving high throughput and energy efficiency. 
Nevertheless, there still has a long way for graph accelerators in practical use for many challenges. As emerging architectural technologies arise, we would also have great opportunities for \hl{to make} significant progresses in not only performance and energy efficiency but also supporting technologies for easy use, evaluation \hl{and} maintenance.

\subsection{Challenges}

\textit{Programmability.}
The development and execution of graph algorithms on existing accelerators rely deeply on the low-level programming with hardware description languages. This enforces that developers have to know the underlying hardware details. Programming for graph programs is non-trivial with a long development cycle. Though high-level programming languages, e.g., C/C++, make this relatively easy, there still lack the effective transformation and mapping of the high-level programming languages to the low-level hardware description languages. The general-purpose \hl{high level synthesis} (HLS) offers a viable solution, which is, however, potentially inefficient due to no full consideration of graph characteristics. It is of great importance to build easy-to-use programming environments for graph processing accelerators.

\textit{Supporting Large Graphs.}
The scale of the graph size is still exploding, which can be easily beyond the available capacity of on-chip memories of a single graph accelerator. For supporting large graphs, an intuitive method is to extend to use larger memory for storing the whole graph. For example, we can use a cluster network of HMCs. However, this may cost a high price at routing the requisite data. 
An alternative approach is to use the heterogeneous graph processing. By using the host memory with more than Terabyte capacity, we can thus have sufficient memory space to store large graphs\upcite{26-FPGP,30-ForeGraph}. Also, a similar design is to connect multiple graph accelerators together and manage them uniformly\upcite{30-ForeGraph,2-EAA}. Nevertheless, the problem is that a significant amount of communication overhead may occur between different graph accelerators. 

\textit{Time-evolving Graphs.}
Existing studies are mostly limited to static graph structures. The graph data may easily change in structure over times. Dynamic graph processing is a hot research topic\upcite{Sha-DynamicGraph, Tornado,SCI-bufferbank}. For example, users of Twitter may update and delete a \hl{post} at anytime. They can also add and delete comments on this \hl{post}. The complex and changeable graph data structure has a high requirement for the latency of graph accelerators. Some methods based on the incremental variation of the subgraph have achieved relatively good results under small scale increments\upcite{Tornado}, but the efficient processing of the large-scale time-evolving graph is still an open problem.

\textit{Complex Attributes of Graphs.}
Different areas have different requirements for the attributes of graphs. For example, two nodes may involve a large number of associated edges that can be handled in parallel. This is common for the server links and road connections. In addition, a number of values can be also associated to a vertex or edge\upcite{zhang-HiddenDimension}. More complex is that the attributes of a graph in \hl{the Graph Network} (GN) can be a vector, set or even another graph\upcite{GN}. These complex attributes of the graphs can result in totally different computing and memory requirements that existing graph processing research can neither fit nor be handled efficiently, let alone hardware circuit designs.

\textit{Machine Learning on Graphs.} Deep learning or machine learning algorithms are also emerging on graphs. There are some research advances on how to represent graph structures into matrics\upcite{graph2vec, structure2vec}. This gives a new dimension of two emerging fields: machine learning and graph processing. 

\textit{Hardware Interfaces.} 
Almost all of existing graph processing accelerators are used solely. They work under the premise that the graph data is placed in its on-chip memory. For supporting large graphs as described previously, requisite external connections to either another accelerators or host processor are needed. This hence requires some extra interfaces for the connection and extension. Unfortunately, few customized graph processing accelerators have such kind of effective interfaces (instead of slow PCI Express lane connection) to support better communication and energy efficiency for graph processing. 

\textit{Tool Chains.}
So far, there are also no convenient tools for programmers to develop and use these graph accelerators easily. Particularly, if the graph programs come across the concurrency and performance bugs, programmers have to rebuild and re-wire the hardware circuit, which is notoriously costly. There still \hl{lacks} of a chain of utility tools for helping understand, diagnose or even fix these low-level problems during development.

\textit{Compiler Support.}
Compiler supporting is an effective way to fill the gap between high-level programming and low-level graph iteration. Symbolic execution is used to parallelize the dependent computations of vertices for achieving compelling performance results on general-purpose processors\upcite{zheng-TACO18}.  Execution parallelism can be also explored for irregular applications by aggressively scheduling execution dependencies at compile time\upcite{AggressivePipeline}. However, more non-trivial efforts are still needed for graph processing accelerators to integrate these advanced compilation features due to the fact that existing (hardware and software) ecosystem surrounding graph accelerators are far from mature.  

\subsection{Opportunities}

\textit{Widespread Adoption.} 
\hl{To the best of our knowledge}, graph processing has been used in many fields, e.g., social network, literature network, traffic network and knowledge atlas. \hl{The} earlier work focuses on addressing typical problems regarding graph searching, random walking and graph clustering. Although there emerge a few latest advances that are attempting to solve the large, complex problems by leveraging graph processing\upcite{zheng-DebugGraph}, \hl{the application of graph processing} still needs to expand. It is a series of open questions regarding how to leverage graph processing and further renovate its hardware acceleration to solve wider practical problems.  

\textit{Emerging Technologies.} 
As discussed before, a few recent studies have used emerging memory technologies (e.g., HMC and ReRAM) to accelerate graph processing, and made the good results in both performance and energy. Nevertheless, the \hl{potentials} of these emerging technologies are still being under-utilized. For instance, GraphR\upcite{10-GRAPHR} uses just one layer ReRAM only, but the fact is that the future ReRAM is often stacked. It is an interesting question on how to use the stacked ReRAM for graph processing acceleration in a more significant way in practice. To this point, more effective and efficient techniques for better supporting emerging technologies have to be settled. 

\textit{FPGA on the Cloud.} FPGAs have been widely adopted in industries to accelerate the datacenter\upcite{Microsoft-Azure} for the high energy efficiency and performance. FPGA providers such as Amazon, 
Baidu, 
and Tencent 
have also offered an easy and flexible programming environment for FPGA development on the cloud. Users can directly program FPGA on the cloud with convenient GUI and sufficient open-source instances\footnote{http://www.plunify.com/en/plunify-cloud/, Jan. 2019.}. The abundant available FPGA resources and integrated development tools provide the opportunities for agile development of FPGA graph processing accelerators\upcite{AGILE}.

\textit{The Rise of Specialized Architectures in Artificial Intelligence.} There has emerged a number of AI specialized \hl{hardware accelerators} in recent years\upcite{TPU, Diannao}. These hardware accelerators have been used to accelerate mechaine learning applications in the cloud\footnote{http://cloud.google.com/tpu/, Jan. 2019.}. The abundant experience of existing AI accelerators can help us understand the underlying architecture design. Besides, a large number of educating resources and developing tools for AI accelerator development can promote the procedure of architecture designs. These opportunities brought by artificial intelligence accelerators can significantly improve the efficiency of graph processing accelerator development.

\section{Conclusions}
With the widely spreading graph applications, \hl{and} gradually increasing data size and the complexity in big data analytics, the performance and energy efficiency of graph processing have brought severe challenges to modern data processing eco-systems. There has emerged a large amount of work that aims at exploring software optimizations to improve the performance and energy efficiency of graph processing under general-purpose architectures, e.g., multi-core CPUs\upcite{zhang-NumaAware} and GPUs\upcite{Medusa, GunRock}.  

However, the significant gap between the unique feature of graph processing and the hardware features of general-purpose architectures limits the further improvement of performance and energy efficiency. Memory access efficiency suffers signficantly \hl{from} traditional memory hierarchy when facing the challenges of the intuitive features in graph processing, e.g., the irregularity and strong dependency\upcite{40-YaoPACT, 1-GRAPHICIONADO}. GPUs also face the drawbacks, e.g., control and memory divergence, load imbalance and global memory access overhead\upcite{Medusa}. That motivates the recent research efforts on developing new hardware architectures for graph processing. 

With the trend and opportunities in domain-specific architectures\upcite{DSA}, e.g., open-source implementations and agile chip development technics\upcite{AGILE}, customized graph processing accelerators have emerged as a promising solution to achieve both high performance and energy efficiency.

In this paper, we investigated a wide spectrum of studies on graph processing accelerators, and \hl{provided} a systematic view \hl{on} their design and implementation. Existing techniques have been categorized into three core aspects: \hl{preprocessing, parallel graph computation and runtime scheduling}. For each aspect, we \hl{reviewed} the state-of-the-art techniques and \hl{made} our remarks on identifying the open problems for future research. 

We also \hl{made} a careful comparison of these studies, and \hl{highlighted} the importance of evaluation benchmarks for graph processing accelerators. At last, we \hl{summarized} the challenges and opportunities of graph processing accelerators, which, we believe, can help architect efficient graph processing accelerators. In summary, graph processing accelerators are still a hot research topic with many technical challenges and opportunities. We call for actions in this survey from different communities, including computer architectures, software systems\hl{, and databases}, to respond these challenges cooperatively. 



\vspace{5mm}

\noindent\parbox{8.3cm}{\parpic{\includegraphics{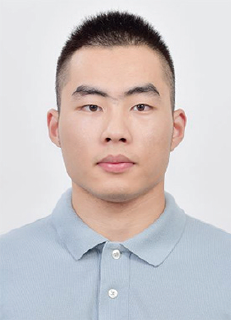}}{\small\quad {\bf Chuang-Yi Gui}  is currently a Ph.D. candidate in the School of Computer Science and Technology at Huazhong University of Science and Technology (HUST), \hl{Wuhan,} China. He received his B.E. degree at HUST in 2017. His current research interests include graph processing and reconfigurable computing.}\\[1mm]}

\noindent\parbox{8.3cm}{\parpic{\includegraphics{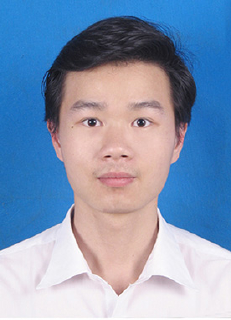}}{\small\quad {\bf Long Zheng}  is now a \hl{postdoctoral researcher} in the school of Computer Science and Technology at Huazhong University of Science and Technology (HUST), \hl{Wuhan,} China. He received his Ph.D. degree at HUST in 2016. His current research interests include program analysis, runtime systems, and configurable computer architecture with a particular focus on graph processing.}\\[1mm]}

\noindent\parbox{8.3cm}{\parpic{\includegraphics{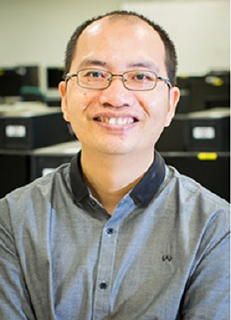}}{\small\quad {\bf Bing-Sheng He}  is currently an Associate Professor at Department of Computer Science, National University of Singapore \hl{(NUS)}. Before joining NUS in May 2016, he held a research position in the System Research group of Microsoft Research Asia (2008-2010) and a faculty position in Nanyang Technological university, Singapore. He got the Bachelor degree in Shanghai Jiao Tong University (1999-2003), and the Ph.D. degree in Hong Kong University of Science \& Technology (2003-2008). His current research interests include Big data management systems (with special interests in cloud computing and emerging hardware systems), Parallel and distributed systems and Cloud Computing.}\\[1mm]}

\noindent\parbox{8.3cm}{\parpic{\includegraphics{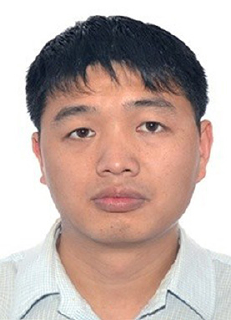}}{\small\quad {\bf Cheng Liu}  is an associate professor of Institute of Computing Technology (ICT), Chinese Academy of Sciences (CAS), Beijing, China. He received his B.E. and M.E. degree in Microelectronic engineering from Harbin Institute of Technology in 2009 and his Ph.D. degree in computer engineering from The University of Hong Kong in 2016. His research focuses on FPGA based reconfigurable computing and domain-specific computing.}\\[1mm]}

\noindent\parbox{8.3cm}{\parpic{\includegraphics{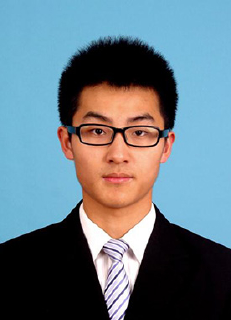}}{\small\quad {\bf Xin-Yu Chen}  is now a Ph.D. candidate of Computer Science in the National University of Singapore, Singapore. He received his B.E. degree in Electronic Science and Technology from Harbin Institute of Technology, Weihai, China, in 2016. His current research interests include FPGA-based heterogeneous computing and database systems.}\\[1mm]}

\noindent\parbox{8.3cm}{\parpic{\includegraphics{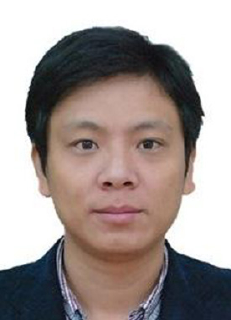}}{\small\quad {\bf Xiao-Fei Liao}  received his Ph.D degree in computer science and engineering from Huazhong University of Science and Technology (HUST), \hl{Wuhan,} China, in 2005. He is now the vice dean in the school of Computer Science and Technology at HUST. He has served as a reviewer for many conferences and journal papers. His research interests are in the areas of system software, P2P system, cluster computing and streaming services. He is a member of the IEEE and the IEEE Computer Society.}\\[1mm]}

\noindent\parbox{8.3cm}{\parpic{\includegraphics{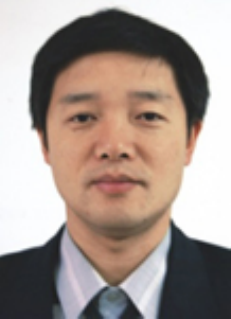}}{\small\quad {\bf Hai Jin}   is a Cheung Kung Scholars Chair Professor of computer science and engineering at Huazhong University of Science and Technology (HUST), \hl{Wuhan}, China. Jin received his PhD in computer engineering from HUST in 1994. In 1996, he was awarded a German Academic Exchange Service fellowship to visit the Technical University of Chemnitz in Germany. Jin worked at The University of Hong Kong between 1998 and 2000, and as a visiting scholar at the University of Southern California between 1999 and 2000. He was awarded Excellent Youth Award from the National Science Foundation of China in 2001. Jin is the chief scientist of ChinaGrid, the largest grid computing project in China, and the chief scientists of National 973 Basic Research Program Project of Virtualization Technology of Computing System, and Cloud Security. Jin is an IEEE Fellow and a member of the ACM. He has co-authored 15 books and published over 600 research papers. His research interests include computer architecture, virtualization technology, cluster computing and cloud computing, peer-to-peer computing, network storage, and network security.}\\[1mm]}

\label{last-page}

\end{multicols}
\label{last-page}
\end{document}